\documentclass[aps,prd,twocolumn,twoside,superscriptaddress,floatfix, nofootinbib]{revtex4-1}
\pdfoutput=1

\usepackage{float}
\usepackage{aas_macros}

\usepackage{color}
\usepackage[utf8]{inputenc} 
\usepackage[T1]{fontenc}    
\usepackage{hyperref}       
\usepackage{url}            
\usepackage{booktabs}       
\usepackage{amsfonts}       
\usepackage{nicefrac}       
\usepackage{microtype}      
\usepackage{lipsum}
\usepackage{graphicx}
\usepackage{amsmath}
\usepackage{comment}
\usepackage{bm}
\mathchardef\mhyphen="2D
\usepackage{soul}

\newcommand{\old}[1]{}
\newcommand{\new}[1]{{#1}}
\newcommand{\newm}[1]{\bm{#1}}

\newcommand{\aktwoh}{A_\mathrm{k2h}} 
\newcommand{\attwoh}{A_\mathrm{t2h}} 
\newcommand{\alk}{\alpha_\mathrm{k}}
\newcommand{\altt}{\alpha_\mathrm{t}}
\newcommand{\xck}{x_\mathrm{c,k}}
\newcommand{\xct}{x_\mathrm{c,t}}
\newcommand{\bk}{\beta_\mathrm{k}}
\newcommand{\bt}{\beta_\mathrm{t}}
\newcommand{\gk}{\gamma_\mathrm{k}}
\newcommand{\gt}{\gamma_\mathrm{t}}

\begin{document}
\title{The Atacama Cosmology Telescope: Modeling the Gas Thermodynamics in BOSS CMASS galaxies from Kinematic and Thermal Sunyaev-Zel'dovich Measurements}

\author{Stefania Amodeo}
\email{stefania.amodeo@astro.unistra.fr}
\affiliation{Department of Astronomy, Cornell University, Ithaca, NY 14853, USA}

\author{Nicholas Battaglia}
\affiliation{Department of Astronomy, Cornell University, Ithaca, NY 14853, USA}
\author{Emmanuel Schaan}
\affiliation{Lawrence Berkeley National Laboratory, One Cyclotron Road, Berkeley, CA 94720, USA}
\affiliation{Berkeley Center for Cosmological Physics, UC Berkeley, CA 94720, USA}
\author{Simone Ferraro}
\affiliation{Lawrence Berkeley National Laboratory, One Cyclotron Road, Berkeley, CA 94720, USA}
\affiliation{Berkeley Center for Cosmological Physics, UC Berkeley, CA 94720, USA}
\author{Emily Moser}
\affiliation{Department of Astronomy, Cornell University, Ithaca, NY 14853, USA}
\author{Simone Aiola}
\affiliation{Center for Computational Astrophysics, Flatiron Institute, New York, NY, USA 10010}
\author{Jason E. Austermann}
\affiliation{Quantum Sensors Group, NIST, 325 Broadway, Boulder, CO 80305}
\author{James A. Beall}
\affiliation{Quantum Sensors Group, NIST, 325 Broadway, Boulder, CO 80305}
\author{Rachel Bean}
\affiliation{Department of Astronomy, Cornell University, Ithaca, NY 14853, USA}
\author{Daniel T. Becker}
\affiliation{Quantum Sensors Group, NIST, 325 Broadway, Boulder, CO 80305}
\author{Richard J. Bond}
\affiliation{Canadian Institute for Theoretical Astrophysics, 60 St. George Street, University of Toronto, Toronto, ON, M5S 3H8, Canada}
\author{Erminia Calabrese}
\affiliation{School of Physics and Astronomy, Cardiff University, The Parade, Cardiff, Wales, UK CF24 3AA}
\author{Victoria Calafut}
\affiliation{Department of Astronomy, Cornell University, Ithaca, NY 14853, USA}
\author{Steve~K.~Choi}
\affiliation{Department of Physics, Cornell University, Ithaca, NY 14853, USA}
\affiliation{Department of Astronomy, Cornell University, Ithaca, NY 14853, USA}
\author{Edward V. Denison}
\affiliation{Quantum Sensors Group, NIST, 325 Broadway, Boulder, CO 80305}
\author{Mark Devlin}
\affiliation{Department of Physics and Astronomy, University of Pennsylvania, 209 South 33rd Street, Philadelphia, PA, USA 19104}
\author{Shannon M. Duff}
\affiliation{Quantum Sensors Group, NIST, 325 Broadway, Boulder, CO 80305}
\author{Adriaan J. Duivenvoorden}
\affiliation{Joseph Henry Laboratories of Physics, Jadwin Hall, Princeton University, Princeton, NJ, USA 08544}
\author{Jo~Dunkley}
\affiliation{Joseph Henry Laboratories of Physics, Jadwin Hall,
Princeton University, Princeton, NJ, USA 08544}
\affiliation{Department of Astrophysical Sciences, Peyton Hall, 
Princeton University, Princeton, NJ USA 08544}
\author{Rolando D\"unner}
\affiliation{Instituto de Astrof\'isica and Centro de Astro-Ingenier\'ia, Facultad de F\'isica, Pontificia Universidad Cat\'olica de Chile, Av. Vicu\~na Mackenna 4860, 7820436, Macul, Santiago, Chile}
\author{Patricio A. Gallardo}
\affiliation{Department of Physics, Cornell University, Ithaca, NY 14853, USA}
\author{Kirsten R. Hall}
\affiliation{Department of Physics and Astronomy, Johns Hopkins University, Baltimore, MD 21218, USA}
\author{Dongwon Han}
\affiliation{Physics and Astronomy Department, Stony Brook University, Stony Brook, NY 11794}
\author{J.~Colin Hill}
\affiliation{Department of Physics, Columbia University, New York, NY, USA 10027}
\affiliation{Center for Computational Astrophysics, Flatiron Institute, New York, NY, USA 10010}
\author{Gene C. Hilton}
\affiliation{Quantum Sensors Group, NIST, 325 Broadway, Boulder, CO 80305}
\author{Matt Hilton}
\affiliation{Astrophysics Research Centre, University of KwaZulu-Natal, Westville Campus, Durban 4041, South Africa}
\affiliation{School of Mathematics, Statistics \& Computer Science, University of KwaZulu-Natal, Westville Campus, Durban 4041, South Africa}
\author{Ren\'ee Hlo\v{z}ek}
\affiliation{David A. Dunlap Department of Astronomy and Astrophysics, University of Toronto, 50 St. George Street, Toronto ON M5S3H4}
\affiliation{Dunlap Institute for Astronomy and Astrophysics, University of Toronto, 50 St. George Street, Toronto ON M5S3H4}
\author{Johannes Hubmayr}
\affiliation{Quantum Sensors Group, NIST, 325 Broadway, Boulder, CO 80305}
\author{Kevin M. Huffenberger}
\affiliation{Department of Physics, Florida State University, Tallahassee, FL 32306, USA}
\author{John P. Hughes}
\affiliation{Department of Physics and Astronomy, Rutgers, the State University of New Jersey, 136 Frelinghuysen Road, Piscataway, NJ 08854-8019, USA}
\author{Brian J. Koopman}
\affiliation{Department of Physics, Yale University, New Haven, CT 06520}
\author{Amanda MacInnis}
\affiliation{Physics and Astronomy Department, Stony Brook University, Stony Brook, NY 11794}
\author{Jeff McMahon}
\affiliation{Kavli Institute for Cosmological Physics, University of Chicago, Chicago, IL 60637, USA}
\affiliation{Department of Astronomy and Astrophysics, University of Chicago, Chicago, IL 60637, USA}
\affiliation{Department of Physics, University of Chicago, Chicago, IL 60637, USA}
\affiliation{Enrico Fermi Institute, University of Chicago, Chicago, IL 60637, USA}
\author{Mathew~S.~Madhavacheril}
\affiliation{Centre for the Universe, Perimeter Institute, Waterloo, ON N2L 2Y5, Canada}
\author{Kavilan Moodley}
\affiliation{Astrophysics Research Centre, University of KwaZulu-Natal, Westville Campus, Durban 4041, South Africa}
\affiliation{School of Mathematics, Statistics \& Computer Science, University of KwaZulu-Natal, Westville Campus, Durban 4041, South Africa}
\author{Tony Mroczkowski}
\affiliation{ESO - European Southern Observatory, Karl-Schwarzschild-Str.\ 2, D-85748 Garching b.\ M\"unchen, Germany}
\author{Sigurd Naess}
\affiliation{Center for Computational Astrophysics, Flatiron Institute, New York, NY, USA 10010}
\author{Federico Nati}
\affiliation{Department of Physics, University of Milano-Bicocca, Piazza della Scienza 3, 20126 Milano, Italy}
\author{Laura B. Newburgh}
\affiliation{Department of Physics, Yale University, 217 Prospect St, New Haven, CT 06511}
\author{Michael D. Niemack}
\affiliation{Department of Physics, Cornell University, Ithaca, NY 14853, USA}
\affiliation{Department of Astronomy, Cornell University, Ithaca, NY 14853, USA}
\author{Lyman A. Page}
\affiliation{Joseph Henry Laboratories of Physics, Jadwin Hall,
Princeton University, Princeton, NJ 08544, USA}
\author{Bruce Partridge}
\affiliation{Department of Physics and Astronomy, Haverford College, Haverford, PA, USA 19041}
\author{Alessandro Schillaci}
\affiliation{Department of Physics, California Institute of Technology, Pasadena, CA91125, USA}
\author{Neelima Sehgal}
\affiliation{Physics and Astronomy Department, Stony Brook University, Stony Brook, NY 11794}
\author{Crist\'obal~Sif\'on}
\affiliation{Instituto de F\'isica, Pontificia Universidad Cat\'olica de Valpara\'iso, Casilla 4059, Valpara\'iso, Chile}
\author{David N. Spergel}
\affiliation{Center for Computational Astrophysics, Flatiron Institute, New York, NY, USA 10010}
\affiliation{Department of Astrophysical Sciences, Peyton Hall, Princeton University, Princeton, NJ, USA 08544}
\author{Suzanne Staggs}
\affiliation{Joseph Henry Laboratories of Physics, Jadwin Hall,
Princeton University, Princeton, NJ, USA 08544}
\author{Emilie R. Storer}
\affiliation{Joseph Henry Laboratories of Physics, Jadwin Hall,
Princeton University, Princeton, NJ, USA 08544}
\author{Joel N. Ullom}
\affiliation{Quantum Sensors Group, NIST, 325 Broadway, Boulder, CO 80305}
\author{Leila R. Vale}
\affiliation{Quantum Sensors Group, NIST, 325 Broadway, Boulder, CO 80305}
\author{Alexander van Engelen}
\affiliation {School of Earth and Space Exploration, Arizona State University, Tempe AZ, 85287, USA}
\author{Jeff Van Lanen}
\affiliation{Quantum Sensors Group, NIST, 325 Broadway, Boulder, CO 80305}
\author{Eve~M. Vavagiakis}
\affiliation{Department of Physics, Cornell University, Ithaca, NY 14853, USA}
\author{Edward J. Wollack}
\affiliation{NASA/Goddard Space Flight Center, Greenbelt, MD, USA 20771}
\author{Zhilei Xu}
\affiliation{Department of Physics and Astronomy, University of Pennsylvania, 209 South 33rd Street, Philadelphia, PA, USA 19104}

\begin{abstract}
The thermal and kinematic Sunyaev-Zel'dovich effects (tSZ, kSZ) probe the thermodynamic properties of the circumgalactic and intracluster medium (CGM and ICM) of galaxies, groups, and clusters, since they are proportional, respectively, to the integrated electron pressure and momentum along the line-of-sight. 
We present constraints on the gas thermodynamics of CMASS (``constant stellar mass'') galaxies in the Baryon Oscillation Spectroscopic Survey (BOSS) using new measurements of the kSZ and tSZ signals obtained in a companion paper [Schaan et al.]. 
Combining kSZ and tSZ measurements, 
we measure within our model the amplitude of energy injection $\epsilon M_\star c^2$, where $M_\star$ is the stellar mass, to be $\epsilon=(40\pm9)\times10^{-6}$,
and the amplitude of the non-thermal pressure profile to be $\alpha_{\rm Nth}<0.2 (2\sigma)$,
indicating that less than 20\% of the total pressure within the virial radius is due to a non-thermal component.
We estimate the effects of including baryons in the modeling of weak-lensing galaxy cross-correlation measurements using the best fit density profile from the kSZ measurement. Our estimate reduces the difference between the original theoretical model and the weak-lensing galaxy cross-correlation measurements in [Leauthaud et al. (2017)] 
by half (50\% at most), but does not fully reconcile it. 
\new{Comparing the tSZ measurements to cosmological simulations, we find that simulations underestimate the CGM pressure at large radii with PTEs ranging from 0.00 to 0.03. The simulations fare better in comparison with the kSZ measurements where we find PTEs ranging from 0.07 to 0.15.} 
This suggests that the energy injected via feedback models in the simulations that we compared against does not sufficiently heat the gas at these radii. We do not find significant disagreement at smaller radii. These measurements provide novel tests of current and future simulations.
This work demonstrates the power of joint, high signal-to-noise kSZ and tSZ observations, upon which future cross-correlation studies will improve.

\end{abstract}

\maketitle

\section{Introduction}
Studying the physical processes and thermodynamic properties that govern the ionized baryons in galaxies and galaxy clusters is essential to our pursuit of understanding galaxy evolution and formation across cosmic time. The circumgalactic medium (CGM) and the intracluster medium (ICM), the baryonic reservoirs for galaxies and clusters, contain the vast majority of baryons in these systems. The thermodynamic properties of the CGM and ICM encode the effects of the assembly history and feedback processes that shape galaxy and cluster formation. Moreover, the impact of baryons and their effects on the underlying dark matter must be known to the percent-level in the non-linear regime if we are to fully utilize the next generation of large-scale-structure cosmological datasets.

The interaction between the CGM or ICM and the photons of the cosmic microwave background (CMB) creates shifts in the photon energy, known as the  ``Sunyaev-Zel'dovich effects'' (SZ \cite{SZ1970, SZ1972}). 
The inverse Compton scattering of the CMB photons with the hot thermal gas causes a
non black-body
distortion in the CMB temperature, known as thermal SZ effect (tSZ), which is proportional to the pressure due to the electrons integrated along the line of sight. The scattering of the CMB photons by the free electrons having bulk motion relative to the CMB rest-frame causes another shift in the CMB temperature, preserving the black-body shape. It is known as kinematic SZ effect (kSZ), and it is proportional to the electron momentum integrated along the line of sight (see \cite{SZreview} for a recent review of the SZ effects).
The analysis of these distortions offers a direct probe of the spatial distribution and abundance of baryons down to the outskirts of galaxies and clusters (see  e.g. \cite{Moodley+09}).  These quantities are still poorly constrained, especially for group-sized structures.
The kSZ effect is particularly well suited to probe low density and low temperature environments like the outskirts of galaxies and clusters, since it is linearly proportional to the electron density and independent of the temperature, and is therefore complementary to the tSZ and X-ray measurements that are more sensitive to the central regions. 
However, while the tSZ has been extensively measured in clusters (and for a wide range of halo masses), kSZ detections (3--5$\sigma$) are relatively new, from stacking analyses \cite{hand+12,Planck_kSZ2015, hill+16,Schaan+16,SPT_DES_kSZ2016,debernardis+17}, or from studies of individual clusters \cite{sayers+13,adam+17,sayers+19}. 
Both the tSZ and kSZ signals contain information about the thermodynamic properties of the CGM and ICM. In a theoretical forecast, \cite{Battaglia+17} showed that combining kSZ and tSZ profile measurements can place tight constraints on baryonic processes like feedback and non-thermal pressure support in the CGM and ICM. Additionally, joint tSZ and kSZ measurements provide constraints on the CGM that are complementary to more traditional probes of the CGM, like absorption line measurements \citep[e.g.,][]{Werk2016,Tumlinson+17,Zahedy2019}

The kSZ measurements \cite{Schaan+16, schaan+20} have traced the distribution of free electrons around galaxies using independent measurements of the peculiar velocity from the galaxy overdensity field. Combining CMB data from the Atacama Cosmology Telescope (ACT) with individual velocity estimates from the CMASS (``constant stellar mass'') catalog of the Baryon Oscillation Spectroscopic Survey  (BOSS  DR10 \cite{BOSSDR10}), they showed that the gas density profiles in groups deviate significantly from a ``dark matter only''  Navarro-Frenk-White profile (NFW \cite{nfw}) expected in absence of feedback. 
This is another manifestation of the well-known ``missing baryon problem'', i.e. that in late-time galaxies and groups, only a small fraction of the cosmological abundance of baryons is found within the virial radius \cite{Fukugita:2004ee, Cen:2006by}. It is speculated that the baryons are pushed out beyond the virial radius by a number of feedback mechanisms, and reside in the outskirts of galaxies in a diffuse and warm state usually referred to as the WHIM (Warm-Hot Intergalactic Medium) \cite{Cen:2006by}. Localizing the ``missing baryons'' by measuring the gas profile out to several times the virial radius is of primary importance for understanding of galaxy formation, interpreting weak lensing measurements, and characterizing the complex physical processes behind feedback.  
While a number of previous observations have made progress in characterizing the missing baryons (see for example \cite{Macquart:2020lln, Kovacs:2018uap, Nicastro:2018eam}), SZ measurements are particularly well-suited to study the outskirts of intermediate and low mass halos, and shed light on this important issue. 

There are observational hints that baryonic effects could be responsible for discrepancies between weak-lensing galaxy cross-correlation measurements and analytic models that do not account for baryons \cite{leauthaud2017}.
On these small scales the baryon distribution no longer traces the dark matter distribution, and thus impacts the matter power spectrum (e.g.  \cite{vD2011,tng2018}). If these baryonic effects are not accounted for in theoretical modeling of the matter power spectrum, then the resulting cosmological parameter inferences will be biased (e.g. \cite{Semboloni2011,Eifler2015}). Understanding the systematic effects from baryons and disentangling them from cosmological information is one of the biggest challenges for the next decade of cosmological surveys, like the Dark Energy Survey  (DES \cite{DES}), the Hyper Suprime-Cam Subaru Strategic Program (HSC-SSP \cite{HSC}), the Vera Rubin Observatory \cite{LSST}, and the Nancy Grace Roman Space Telescope \cite{Roman}. Theoretical models exist that are calibrated using observations of the baryon fraction \cite{Schneider2019} or simulations \cite{vD2020,Mead2020}. In this paper, we use 
kSZ measurements of the gas profile that probe the baryon distribution on smaller scales and we explore a simple empirical model of this relationship.

Large cosmological simulations provide a partially predictive model and are now able to reproduce many of the optical properties of galaxies (e.g. \cite{fire,Eagles,SomervilleDave2015,HorizonAGN,tng2018}). It is necessary for these simulations to include physically motivated {\it sub-grid} modeling schemes for processes like star formation and various energetic feedback mechanisms, since they do not resolve the scales necessary to perform {\it ab initio} calculations of these critical physical processes in galaxy evolution. With kSZ and tSZ cross-correlation measurements we can directly test and inform these sub-grid feedback models \cite{Battaglia+17,BH2019,DSR}, as these sub-grid models are not precisely tuned to reproduce SZ observations of the CGM and ICM. 

In this paper, we use the new stacked tSZ and kSZ measurements obtained in a companion paper \cite{schaan+20} by cross-correlating the CMASS galaxy catalogs of BOSS DR10 \cite{BOSSDR10} and temperature maps from combined ACT DR5 and \emph{Planck} data from \cite{naess+20} in the f150 and f090 bands (centered at roughly 150 GHz and 98 GHz,  respectively), to constrain the baryonic processes, such as feedback and non-thermal pressure support, and the thermodynamic profiles of the CGM/ICM.
Our models include a correction for the contamination of the tSZ signal by the thermal emission of dust from galaxies in our sample, which we estimate from ACT DR5 (f090, f150) and Herschel/H-ATLAS data \cite{eales+10} in the three bands centered at 500 $\mu$m (600 GHz), 350 $\mu$m (857 GHz), and 250 $\mu$m (1200 GHz). We compare our results to predictions from Illustris TNG cosmological simulations \cite{tng2018} and older simulations by \cite{BBPSS2010}. Finally, we use the best-fit density profiles to estimate the effects of including baryons in the modeling of weak-lensing galaxy cross-correlation measurements by \cite{leauthaud2017}.

Two upcoming papers \cite{Vavagiakis+20,calafut+20}, present kSZ (5$\sigma$) and tSZ (${\sim}10\sigma$) measurements using the same ACT and \emph{Planck} maps as those used here,
but different galaxy samples from BOSS.
They explore the luminosity dependence of the signals, as well as the shape of the velocity correlation function. These probes contain information about dark energy and modifications to General Relativity \cite{Mueller:2014nsa}, neutrino masses \cite{Mueller:2014dba} and primordial non-Gaussianity \cite{munchmeyer+19}. Because the galaxy samples are different, with different host halo masses, the results from these two papers are not directly comparable to ours.

In this paper and \cite{schaan+20}, our main interest is instead in the radial dependence of the kSZ and tSZ signals, and particularly the baryon profiles.
We focus on the CMASS galaxy sample, for which clustering and galaxy lensing measurements are available, in order to obtain a complete picture of the gas thermodynamics.
This allows us to constrain the properties of feedback in these halos, and shed new light on the low lensing tension \cite{leauthaud2017}.
Because these two pairs of papers focus on different information from the kSZ and tSZ signals, they use different estimators: 
the pairwise kSZ estimator in \cite{Vavagiakis+20,calafut+20} is particularly suited to measure the velocity correlation function,
whereas the stacking with reconstructed velocities of \cite{schaan+20} and this work is convenient for measuring the baryon profiles.
Overall, these two pairs of papers are complementary, and highlight the wealth of information in joint kSZ and tSZ measurements.

The paper is organized as follows: 
Section \ref{sec:modelobs} describes our modeling of the tSZ and kSZ signals in terms of both a polytropic gas model that includes energetic feedback and a non-thermal pressure component, and parametric (generalized NFW) models for the gas thermal pressure and density. We give in Appendix \ref{sec:twohalo} details on how we account for the contribution to the halo gas profiles from neighboring halos (i.e. two-halo term).
The procedure described in this section is implemented in the publicly available code \texttt{Mop-c GT} (``Model-to-observable projection code for Galaxy Thermodynamics'') \footnote{\url{https://github.com/samodeo/Mop-c-GT}}.
We present the constraints on our models, using the kSZ and tSZ profiles measured by \cite{schaan+20}, in Section \ref{sec:results}, and provide details on the dust correction in Appendix \ref{sec:dust}. In Section \ref{sec:lensing} we estimate the impact of baryons on galaxy-galaxy lensing measurements of the CMASS sample and compare our results to current observations.
In Section \ref{sec:sims}, we compare the kSZ/tSZ observations to predictions for the density and pressure from hydrodynamical simulations by running the simulations through the same projection code.
We summarize our results and draw conclusions in Section \ref{sec:summary}.  

We adopt a flat $\Lambda$CDM cosmology with matter density $\Omega_m=0.25$,  baryon  density $\Omega_b=0.044$, dark  energy density $\Omega_\Lambda=0.75$, and local expansion rate $H_0=70$ km s$^{-1}$ Mpc$^{-1}$ ($h\equiv H_0/$($100$ km s$^{-1}$ Mpc$^{-1}$). The choice of cosmological parameters does not significantly affect our results. Halo masses are quoted as $M_{\rm 200}$, at a radius of $R_{\rm 200}$, within which the halo density is $200$ times the critical density of the universe at the halo’s redshift, $\rho_{\rm cr}(z) \equiv 3H_0^2 (\Omega_m (1+z)^3+\Omega_\Lambda)/(8\pi G)$.

\section{Modeling the observed signal}
\label{sec:modelobs}
We describe the kSZ and tSZ radial profile data and their relationship to the gas density and thermal pressure profiles in Section \ref{sec:modelobs_formulae}.
We parametrize the three-dimensional profiles of these quantities using two halo models presented below: the Ostriker-Bode-Babul, ``OBB'' \cite{Ostriker+05}, and the generalized Navarro-Frenk-White, ``GNFW'' \cite{Zhao1996,Battaglia+12b}.
In section \ref{sec:OBBmodel}, we describe our use of the OBB model and describe how it provides constraints on the non-thermal pressure profile and the star formation feedback process. In section \ref{sec:GNFWmodel}, we describe our use of the GNFW model, while in Section \ref{sec:modelobs_twohalo}, we describe how we handle two-halo effects.

\subsection{kSZ and tSZ effects}
\label{sec:modelobs_formulae}
We use stacked CMB temperature measurements from \cite{schaan+20} obtained by cross-correlating combined ACT DR5 and \emph{Planck} temperature maps from \cite{naess+20} in the f090 and f150 bands, with the CMASS spectroscopic catalog of galaxies \cite{BOSSDR10} in the region covered by ACT. 
The CMASS galaxies span a redshift range $0.4<z<0.7$, with a median redshift $z=0.55$. Approximately 85\% of them reside at the center of galaxy groups or clusters with mean stellar mass $M_{\rm \ast} = 3\times10^{11} M_\odot$ (from \cite{maraston+13} stellar mass estimates), corresponding to a
halo mass $M_{\rm halo} \sim 3\times10^{13} M_\odot$, according to the stellar--halo mass conversion of \cite{KVM}.

The tSZ and kSZ signals are measured from microwave temperature maps by applying a compensated aperture photometry (CAP) filter at the position of each galaxy; we average the value of the pixels within a disk of radius $\theta_{d}$ and subtract the average of the pixels in an adjacent, equal area ring with external radius ${\sqrt 2}\theta_{d}$. With the ACT CMB maps in temperature units relative to the CMB ($\mu$K), the output of the CAP filter is given by: 
\begin{equation}
\label{eq:ap}
AP(\theta_{d})=
\int d^2\theta \, \delta T(\theta) \, W_{\theta_{d}}(\theta) \,,
\end{equation}
with units of $\mu K\cdot\text{arcmin}^2$, where the angular CAP filter function $W_{\theta_{d}}(\theta)$ is dimensionless, defined as:
\begin{equation}
W_{\theta_{d}}(\theta) =
\left\{
\begin{aligned}
1& &  &\text{for} \, \theta < \theta_d \,, \\
-1& &  &\text{for} \, \theta_d \leq \theta \leq \sqrt{2}\theta_d \,, \\
0& & &\text{otherwise}. \\
\end{aligned}
\right.
\end{equation}
The filter aperture $\theta_{d}$ has been chosen to vary between 1 and 6 arcmin, corresponding to approximately 1--4 times the typical virial radius $R_{\rm vir}$,
in order to investigate the physical scales relevant for the effects of feedback. 
The tSZ and kSZ signals are measured with a signal-to-noise ratio of 11 and 8, respectively \cite{schaan+20}, from ACT+\emph{Planck} coadded maps in two frequency bands, f090 and f150 \cite{naess+20}. In order to properly model this specific set of data, we convolve 2D-projected temperature profiles to the same beams with which the \cite{naess+20} ACT+\emph{Planck} coadded maps are convolved. These beams have non-Gaussian, scale-dependent profiles, with full-widths at half-maximum of 2.1 (f090) and 1.3 (f150) arcmin (see Fig.~15 in \cite{schaan+20}).
The tSZ and kSZ signals can be modelled in terms of temperature fluctuations as described below. 

The tSZ temperature fluctuations are given by:
\begin{equation}
\label{eq:tsz}
    \frac{\Delta T_{\rm{tSZ}}}{T_{\rm CMB}} = f(\nu) y \, ,
\end{equation}
where the frequency dependence, neglecting relativistic corrections (e.g. \cite{Nozawa+06, Chluba+12}), is given by $f(\nu) = x \coth{(x/2)} -4$, with $x=h\nu / k_B T_{\rm CMB}$,  $T_{\rm CMB}$ is the CMB temperature, $k_{\rm B}$ is the Boltzmann constant, and the Compton-$y$ parameter measured within $\theta$, at an angular diameter distance to redshift $z$, $d_A(z)$, is:
\begin{equation}
\label{eq:y}
    y(\theta) = \frac{\sigma_T}{m_e c^2} \int_{los} P_e (\sqrt{l^2 + d_A(z)^2 |\theta|^2}) ~dl \, .
\end{equation}
Here $\sigma_T$ is the Thomson cross-section, $m_e$ is the electron mass, $c$ is the speed of light, $P_e$ is the thermal electron pressure and $dl$ is the line-of-sight ($los$) physical distance.

The kSZ temperature fluctuations are given by:
\begin{equation}
\label{eq:ksz0}
    \frac{\Delta T_{\rm{kSZ}}}{T_{\rm CMB}} =  \frac{\sigma_T}{c} \int_{los} e^{-\tau} n_e  ~v_p ~dl \, ,
\end{equation}
where $n_e$ is the electron number density, $v_p$ is the peculiar velocity and $\tau$ is the optical depth to Thomson scattering along the line of sight, defined as:
\begin{equation}
\label{eq:tau}
    \tau(\theta) = \sigma_T \int_{los} n_e (\sqrt{l^2 + d_A(z)^2 |\theta|^2}) ~dl \, .
\end{equation}
The mean optical depth in our redshift range ($0.4<z<0.7$) is below one percent (see e.g. \cite{planck2016_reion}), therefore we approximate the $e^{-\tau}$ factor in the integral as 1. Moreover, since \cite{schaan+20} selectively extract the kSZ signal correlated with the galaxy group of interest, we can further simplify Equation\ref{eq:ksz0} as:
\begin{equation}
\label{eq:ksz}
    \frac{\Delta T_{\rm{kSZ}}}{T_{\rm CMB}} =  \tau_{\rm  gal} \left( \frac{v_r}{c} \right)  \, ,
\end{equation}
where $\tau_{\rm  gal}$ refers to the optical depth of the galaxy group considered, and $v_r=1.06\times10^{-3} c$ is the RMS of the peculiar velocities, projected along the line of sight, where the magnitude adopted is for the median redshift of the CMASS sample, $z = 0.55$, in the linear approximation.
We estimate the uncertainty on the velocity reconstruction being less than few percent \cite{schaan+20} and given our  current signal-to-noise, we do not propagate it into the uncertainty on the kSZ profile. However, this will be crucial for upcoming measurements with higher kSZ signal-to-noise ratio \cite{Nguyen+20}.

The electron density and pressure can be converted into the gas density $\rho_{\rm gas}$ and thermal pressure $P_{\rm th}$. Assuming a fully ionized medium with primordial abundances:
\begin{equation}
\label{eq:el_prof}
\begin{aligned}
n_e &= \frac{(X_{\rm H} +1)}{2} \frac{\rho_{\rm gas}}{m_{\rm amu}} \, , \\
P_e &= \left(\frac{2+2X_{\rm H}}{3+5X_{\rm H}} \right) P_{\rm th} \,,
\end{aligned}
\end{equation}
where $X_{\rm H} = 0.76$ is the hydrogen mass fraction,  $m_{\rm amu}$ is the atomic mass unit.

Therefore, the tSZ and kSZ temperature fluctuations are related to the gas thermal pressure $P_{\rm{th}}$ and to the gas density $\rho_{\rm{gas}}$, respectively. 

In order to model the observed signal we apply the same aperture photometry filters used in the analysis of the observations by substituting the temperature models for the kSZ and tSZ (Eq.\ \ref{eq:tsz}-\ref{eq:y} and \ref{eq:ksz}-\ref{eq:tau}, respectively) in Eq.\ \ref{eq:ap}. 

To summarize, \emph{i}) we project the 3D gas profiles along the line of sight as in Eqs. \ref{eq:y},\ref{eq:tau}, \emph{ii}) we
convolve them with the beam profile measured at f090 and f150, \emph{iii}) for the pressure model we also multiply by the map response to the tSZ in each band \cite{naess+20}, \emph{iv}) we then get the average temperature within disks of varying radii $\theta_{d}$, \emph{v)} for each aperture, we subtract the mean temperature in an adjacent ring of external radius $\sqrt{2}\theta_{d}$ and equal area, so as to reproduce the same aperture photometry filtering applied to data.

These profiles from both OBB and GNFW are defined for a halo of given mass and redshift. We compute average profiles that account for the mass distribution of the CMASS sample. Using mass-weighted averages is particularly important for the tSZ modeling, as the tSZ signal is proportional to $M^{5/3}$, while the kSZ is linearly dependent on mass. We do not average over the distribution of redshifts, which is peaked around the median (see Fig.\ 2  in \cite{schaan+20}), and we use therefore the median redshift of our sample, z=0.55. Using test models, we have checked that computing mass and redshift-weighted average profiles does not significantly change the results, for both density and pressure. Our modeling of the CMASS sample does assume that all CMASS galaxies are central galaxies, which is reasonable given the CMASS selection and our current measurement errors. Additionally, we assume the direct mapping between stellar mass and halo mass used on the CMASS sample \cite{KVM}.

\subsection{OBB model}
\label{sec:OBBmodel}
In order to investigate the thermodynamic properties of the CGM and ICM, we implement a model proposed by \cite{Ostriker+05} (see also \cite{Bode+09, Shaw+10}). 
The model assumes that the gas has an initial energy per unit mass equivalent to that of the dark matter halo. In our implementation, we assume that the dark matter follows a spherically symmetric NFW density profile \cite{nfw}, characterized by a density normalization $\rho_0$ and a scale radius $r_s$:
\begin{equation}
\label{eq:nfw}
    \rho_{\rm DM}(x) = \frac{\rho_0}{x(1+x)^2} \,,
\end{equation}
where $x\equiv r/r_s$. The scale radius is related to the halo mass through the concentration parameter $c_{\rm NFW}$, $r_s = R_{\rm 200}/c_{\rm NFW}$. We use the concentration-mass power-law relation by \cite{Duffy+08} obtained from N-body simulations for halo masses in the range $10^{11}-10^{15}h^{-1} M_\odot$ at $0<z<2$: 
\begin{equation}
    c_{\rm NFW} = 5.71 \times (1 + z)^{-0.47} \times \left(\frac{M}{2\times10^{12} M_\odot}\right)^{-0.084} \,,
\end{equation} with a scatter of 0.15 dex.
The assumption of spherical symmetry for the sample should be accurate for both the gas and the dark matter since we are modeling stacked profiles (e.g.~\cite{CorlessKing2007,BeckerKravtsov2011, Battaglia+12a}).
Assuming that the initial gas mass is a fraction of the total halo mass equal to the cosmic baryon fraction ($M_{\rm gas,i} = \Omega_b/\Omega_m M_{\rm tot,i}$), we can use the virial theorem to get the gas energy and surface pressure in terms of the dark matter halo parameters. As the system evolves, some fraction of the initial ICM gas will cool and turn into stars, lowering the gas mass and increasing the energy per unit mass of the remaining gas \cite{VoitBryan2001}, some work $\Delta E_p$ will be done by the surface pressure for changes of the gas volume, and some energy can be injected into the gas by feedback processes from supernovae and active galactic nuclei (AGN). Finally, the model assumes that the gas rearranges itself into a polytropic distribution characterized by a central pressure $P_0$ and density $\rho_0$. The final ICM density $\rho_{\rm gas}(r)$ and pressure $P_{\rm tot}(r)$ profiles are given by:
\begin{equation} \label{eq:eq_obb}
    \begin{aligned}
    \rho_{\rm gas}(r) &= \rho_0 \theta(r)^{\frac{1}{\Gamma-1}} \,,\\
    P_{\rm tot}(r) &= P_0\theta(r)^{\frac{1}{\Gamma-1}+1} \,,
    \end{aligned}
\end{equation}
where $\Gamma$ is the polytropic index and $\theta(r)$ is the polytropic variable, defined as:
\begin{equation}
    \label{eq:polyfunc}
    \theta(r) = 1+ \frac{\Gamma-1}{\Gamma}\frac{\rho_0}{P_0} \left(\Phi_0 - \Phi(r) \right) \,.
\end{equation}
Here $\Phi_0$ is the central gravitational potential of the halo and the system is in equilibrium: $dP_{\rm tot}/dr = -\rho_{\rm gas} d\Phi/dr$.
The total pressure includes a non-thermal component that is mainly attributed to bulk motions and turbulence caused by gas accretion and/or merging structures. Hydrodynamical simulations have reported that these processes contribute up to 10-30\% of the total pressure and that this amount increases with radius (see e.g. \cite{NormanBryan1999,Rasia+04,Faltenbacher+05,Rasia+06,Lau+09, Nelson+14, Shi+15}). 
Following the model of \cite{Shaw+10}, we describe the radial  profile of non-thermal pressure component as a power law of the form:
\begin{equation}
    P_{\rm Nth}(r) = \alpha_{\rm Nth} (r/R_{\rm 200})^{n_{\rm Nth}} \, P_{\rm tot}(r)   \,.
\end{equation}
We fix the radial dependence to $n_{\rm Nth}=0.8$ as in \cite{Shaw+10}. That is the best-fit value to the hydrodynamical simulations of \cite{nagai+07}, where the non-thermal pressure is measured from the radial velocity dispersion of the gas in shells of increasing radii. 
We leave the normalization $\alpha_{\rm Nth}$ as a free parameter in the fit.
This non-thermal pressure model has been subsequently validated by other simulations \cite{Battaglia+12a, Nelson+14}.
The thermal pressure is then: 
\begin{equation}
\label{eq:pth_eq_obb}
P_{\rm th}(r) = P_{\rm tot}(r)-P_{\rm Nth}(r) \,.    
\end{equation}
 
The gas distribution can finally be solved for $P_0$ and $\rho_0$ by imposing the conservation of energy and defining the boundary condition: the final gas energy $E_f$ will be equal to the initial energy $E_i$ plus the energy injected by feedback processes, $\epsilon M_\star c^2$, where $\epsilon$ is a dimensionless parameter quantifying the efficiency of the feedback and $M_\star$ is the stellar mass, plus the energy $\Delta E_p$ due to expansion or contraction of the halo boundaries:
\begin{equation}
    E_f = E_i + \epsilon M_\star c^2 + \Delta E_p \,,
\end{equation}
with the condition that the total pressure at the halo boundary must match the initial surface pressure.

This model was found to be in good agreement with high-resolution hydrodynamic simulations and can reproduce the observed X-ray scaling relations. For massive clusters ($M_{\rm tot}=10^{14} h^{-1} M_\odot$), \cite{Ostriker+05}  find a good fit to \old{the observations} simulations with a polytropic index $\Gamma=1.2$, a fraction of the baryonic mass condensed into stars that is transferred back to the remaining gas, estimating the feedback efficiency, of $\epsilon = 3.9\times10^{-5}$, and 10\% of the total pressure due to a non-thermal component.
For clusters in a similar mass range, \cite{Bode+09} give a comparable amount of feedback, between $3$ and $5\times10^{-5}$, while \cite{Shaw+10} adopt a smaller value in their fiducial model, $\epsilon = 10^{-6}$, and a redshift-dependent non-thermal pressure parameter in the range  $\alpha_{\rm Nth}=[0-0.33]$.
We implement this model to constrain the normalization of non-thermal pressure profile, $\alpha_{\rm Nth}$ and the energy injected in the gas by feedback, $\epsilon$.

\subsection{GNFW models}
\label{sec:GNFWmodel}

We parametrize the three-dimensional profiles of the gas density and  the thermal pressure, respectively, using two generalized NFW models. For both models, we choose parameters and parameter ranges motivated by fits to cosmological simulations described below. 
In general, our SZ measurements constrain the shape of the pressure and density profiles at large radii, while they do not constrain that well the parameters which are sensitive to the profile properties at small radii. Thus, we fix the values of such parameters motivated by the simulations mentioned below when we fit the GNFW models. 

For the density model, 
we refer to the following generalized NFW profile \cite{Zhao1996}: 
\begin{equation}
\label{eq:rho_gnfw}
\begin{aligned}
    \rho_{{\rm GNFW}}(x) &= \rho_0 (x/\xck)^{\gk} [1+(x/\xck)^{\alk}]^{-\frac{\bk-\gk}{\alk}} \,,\\
    \rho_{\rm gas}(x) &= \rho_{{\rm GNFW}}(x) ~\rho_{\rm cr}(z) ~f_b \,,
\end{aligned}
\end{equation}
where $x \equiv r/R_{200}$, $\xck$ is a core scale, ($\alk$,$\bk$,$\gk$) are the slopes at $x\sim1$, $x\gg1$, and $x\ll1$, respectively, $\rho_{\rm cr}(z)$ is the critical density of the Universe at redshift $z$, and $f_b = \Omega_b/\Omega_m$ is the baryon fraction. 
Given the considerable degeneracy of the parameters we fix two parameters that are sensitive to the profile properties at small radii, $\gk = - 0.2$, and $\alk = 1$, as in \cite{Battaglia+16}.

For the thermal pressure model,
we use a slightly modified GNFW profile following \cite{Battaglia+12b}:
\begin{equation}
\label{eq:P_gnfw}
\begin{aligned}
P_{{\rm GNFW}}(x) &= P_0 (x/\xct)^{\gt} [1+(x/\xct)^{\altt}]^{-\bt}\,, \\
P_{\rm th}(x) &= P_{{\rm GNFW}}(x) P_{\rm 200}\,,
\end{aligned}
\end{equation}
where $P_{\rm 200} = G M_{\rm 200}~200~\rho_{\rm cr}(z)~f_b/(2R_{\rm 200})$, $\xct$ is a core scale, ($\altt$,$\bt$,$\gt$) are the slopes at $x\sim1$, $x\gg1$, and $x\ll1$, respectively. These parameters define the pressure radial profile and are different from the parameters in Eq.~\ref{eq:rho_gnfw}. 
There is significant degeneracy among all the GNFW parameters, so we fix two parameters sensitive to the profile properties at small radii, at the values suggested by previous cosmological simulations: $\gt = -0.3$ \cite{nagai+07, Battaglia+12b} and $\xct = 0.497 \times (M_{\rm 200}/10^{14} M_\odot)^{-0.00865} \times (1+z)^{0.731}$ \cite{Battaglia+12b}. 

\subsection{Two-halo term}
\label{sec:modelobs_twohalo}
For both profiles we include a two-halo term. The total density and pressure profiles are modeled as $\rho(r) = \rho_{\rm one\mhyphen halo}(r) + \aktwoh \,\rho_{\rm two\mhyphen halo}(r)$ and $P(r) = P_{\rm one\mhyphen halo}(r) + \attwoh \, P_{\rm two\mhyphen halo}(r)$, respectively, where the one-halo terms are computed according to the models above. In Appendix~\ref{sec:twohalo} we show how the fiducial profiles for the two-halo terms are calculated and we include free parameters, $ A_{\rm 2h}$ in front of those terms for both density and pressure that scale the amplitude and include them in our fits. 

\section{Results}
\label{sec:results}
Ref.~\cite{schaan+20} presents the results for the individual integrated aperture quantities (the
optical depth to Thomson scattering and the Compton y), which is a standard practice for SZ cross-correlation measurements. 
Here we study instead how these SZ cross-correlations (the thermodynamic profiles) change as a function of the distance from the galaxy center. Given the good S/N of the \cite{schaan+20} measurements in each radial aperture, we are able to move beyond single aperture analyses and use the information from all the scales we can access in these measurements. We can thus improve our ability to constrain models or simulations of the CGM, while this constraining power would be reduced if we compressed the SZ measurements into a single aperture. 

In this section, we present constraints on feedback and non-thermal pressure from combined kSZ and tSZ profile data (Sec. \ref{sec:OBB}) and constraints on our parametric models of the gas density (Sec. \ref{sec:gnfw_density}) and thermal pressure profile (Sec. \ref{sec:gnfw_pressure}). We begin each subsection by describing the free parameters in the fit and motivating the use of priors, then we show the results of fitting in Tables \ref{tab:mcmc_OBB_par} and \ref{tab:gnfw_par} and Figures \ref{fig:OBBmcmc} to \ref{fig:GNFW_tSZ_mcmc}.  

For our fits, we use the data ($\vec{d}$) and covariance matrices ($\mathbf{C}$) estimated by \cite{schaan+20} using the bootstrap method. 
We use all the data points between 1 and 6 arcmin when we fit the kSZ profile, while we exclude the smallest aperture when we fit the tSZ profile since the dust contamination is a large fraction of the signal there.

We assume the likelihood ($\mathcal{L}$) to be Gaussian, written as:
\begin{equation}
    \ln \mathcal{L}[\vec{d}|\vec{m}(\vec{\theta})] = - \frac{1}{2} \left[ \vec{d}-\vec{m}(\vec{\theta}) \right]^{\rm T}   \mathbf{C}^{-1} \left[ \vec{d}-\vec{m}(\vec{\theta}) \right] \,,
\end{equation}
where $\vec{m}(\vec{\theta})$ is the model evaluated at the parameter $\vec{\theta}$. 
The posterior on the model parameters ($\mathcal{P}$) is then expressed as:
\begin{equation}
  \ln \mathcal{P}(\vec{\theta}|\vec{d}) = \ln [ \mathcal{L}(\vec{d}|\vec{m}(\vec{\theta})) \mathrm{Pr}(\vec{\theta})]   \,,
\end{equation}
where $\mathrm{Pr}(\vec{\theta})$ are the priors on $\vec{\theta}$.
We use Markov chain Monte Carlo calculations (MCMC \cite{Metropolis,Hast70}) to estimate the posterior probability functions, with the Affine-Invariant Ensemble Sampler algorithm implemented in \texttt{emcee} \cite{emcee}. We run multiple \texttt{emcee} ensembles, adding independent sets of chains until the Gelman-Rubin convergence parameter, $\hat R$, reaches values smaller than 1.1 \cite{GelmanRubin}.

\subsection{OBB model}
\label{sec:OBB}

We probe the efficiency of energetic feedback and non-thermal pressure in the CMASS sample by fitting the OBB model, described in Sec.\ \ref{sec:OBBmodel}, to kSZ and tSZ data simultaneously.
The tSZ likelihood includes the dust correction described in Appendix \ref{sec:dust}.
We assume a uniform prior for the polytropic index in the range $1<\Gamma<5/3$. The lower limit guarantees the existence of the polytropic function (Eq.\ \ref{eq:polyfunc}), while the upper limit excludes non-relativistic degenerate gas. 
For the other parameters we assume uniform priors within physically reasonable ranges: $0.01<\alpha_{\rm Nth}<0.8$, $-4.8<\log_{\rm 10} \epsilon<-4.0$,  $0.1<A_{\rm k2h}<5$,  $0.1<A_{\rm t2h}<5$. We also sample the dust parameters and we marginalize over them as discussed in Appendix \ref{sec:dust}.

Figure \ref{fig:OBBmcmc} shows the posterior contours of the OBB parameters and the 2$\sigma$ range of the OBB model parameters derived from the kSZ and tSZ data. Our best-fit model is shown with solid lines.
We obtain $\chi^2=86.6$ with a probability to exceed this value (PTE) of 0.15 for the joint kSZ+tSZ(+dust) fit, indicating a good description of the data. We compute the PTE values from 10000 Monte Carlo random samples with mean zero and using the covariances of our measurements.
This is a better estimator than the reduced $\chi^2$ since the number of degrees of freedom is unknown for non-linear models \cite{andrae+10}. 
Table \ref{tab:mcmc_OBB_par} reports the marginalized constraints ($1\sigma$). We find a non-zero amount of feedback from AGN and supernovae, $\epsilon=(40\pm9)\times10^{-6}$, 
that we robustly estimate with a 23\% precision within the context of our model. 
From this result we can estimate that the fraction of energy injected into CMASS galaxies is $E_{inj} = \epsilon M_\star c^2 \simeq (2.2^\pm0.5)\times10^{61} \text{erg}$, using their mean stellar mass $M_{\rm \ast} = 3\times10^{11} M_\odot$. Compared to the total binding energy for these halos $U = (3/5) ~G (4/3\pi)^{1/3} (200 \rho_c)^{1/3} M_{\rm halo}^{5/3} \simeq 7.2\times10^{61} \text{erg}$, using a mean total of $M_{\rm halo} = 3\times10^{13} M_\odot$, the energy injected by feedback is 30\% (23-37\% given our 1$\sigma$ error bars on the $\epsilon$ parameter). This is a lower limit to the energy injected as we have not included radiative losses from cooling. Also we have not propagated any uncertainties on the stellar or total masses of the CMASS sample. The ratio of energy injected to the binding energy that we estimate is consistent with the 25\% value that can be calculated from the results in \cite{BBPSS2010}.

Ref.~\cite{flender+17} uses X-ray measurements of gas density and mass in galaxy groups and clusters to calibrate a model similar to the model used in this work.  
They find a feedback efficiency factor of $\epsilon=(4_{-3}^{+5})\times10^{-6}$ (95\% confidence level) that is smaller, at $\sim4\sigma$, than our result. However, we note that their X-ray sample is on average more massive than ours and the parameter space explored is also different. 
We find an upper limit for the amplitude of the non-thermal pressure profile,
$\alpha_{\rm Nth}<0.2 (2\sigma)$ indicating that less than 20\% of the total pressure within $R_{\rm 200}$ is due to a non-thermal component.
This amount is comparable to the upper limit of 33\% found by \cite{Shaw+10} and it is also consistent with the non-thermal pressure fraction constrained by \cite{Shi+15} for massive clusters at $z=0$, from the turbulence-to-total ratio of the cluster velocity dispersion.
\old{The levels of uncertainty are consistent with the constraints on the same model forecasted by Battaglia+17 using a Fisher analysis, who find 12\% on $\epsilon$ and 24\% on $\alpha_{\rm Nth}$ for a Stage-3 CMB$+$CMASS survey.}

Using the posterior distributions of the fit parameters we obtain profiles of the gas density and thermal pressure from Eqs.\ \ref{eq:eq_obb}-\ref{eq:pth_eq_obb}, and estimate the electron temperature profile as $\rm{T_e = P_e / (n_e k_{\rm B})}$, where we calculate $\rm{P_e}$ and $\rm{n_e}$ from Eq.\ \ref{eq:el_prof}. 
Figure \ref{fig:Te_prof} shows the median (black line) and the 2$\sigma$ range (blue band) of the models obtained from the MCMC chains. 
\old{We constrain the temperature profile at more than $6\sigma$ in the first two radial bins, i.e. within approximately the virial radius, and better than $3\sigma$ within $\sim2R_{\rm vir}$. We still have $2\sigma$ constraints at $\sim3R_{\rm vir}$, but less than $2\sigma$ at larger radii. We find a decreasing temperature profile from $(8.3\pm0.9)\times10^6$~K ($0.70\pm0.08$~keV) to $(1.0\pm1.8)\times10^6$~K ($0.09\pm0.15$~keV) in our radial range.} 
We constrain the temperature profile at more than $4\sigma$ in the first two radial bins, i.e. within approximately the virial radius, at better than $2\sigma$ within $\sim2R_{\rm vir}$ and less than $2\sigma$ at larger radii. We find a decreasing temperature profile from $(1.3\pm0.2)\times10^7$~K ($1.1\pm0.1$~keV) to $(1.8\pm2.4)\times10^6$~K ($0.2\pm0.2$~keV) in our radial range.
These values are overall consistent with the mean electron temperature estimated by \cite{schaan+20} as the ratio of tSZ and kSZ measurements for each aperture photometry radius. For reference, we compute the expected virial temperature as $T_{\rm vir}=\mu m_{\rm p} G M_{\rm 200}/2 k_{\rm B} R_{200}$ \cite{MVW2010}, assuming a singular isothermal sphere of gas of mass equal to the the mean mass of our CMASS sample ($M_{\rm 200}=3.3\times10^{13}  M_\odot$), where $\mu\approx1.14$ is the mean molecular weight for a fully ionized medium with primordial abundances. We get $T_{\rm vir}=1.7\times10^7$ K, that is of the same order of magnitude of our measured profile.   
In a recent paper, \cite{Lim+20} compare gas profiles and temperatures obtained from \emph{Planck} SZ cross-correlation measurements of halos in the Sloan Digital Sky Survey (SDSS \cite{sdss}) at $z\sim0$, with different cosmological simulations.  We note that our temperature measurements are higher, although not directly comparable because the samples are different and the scales involved in the analysis are different, mostly dictated by the different resolution of \emph{Planck} and ACT.

\begin{figure*}
\centering
  \includegraphics[width=\textwidth]{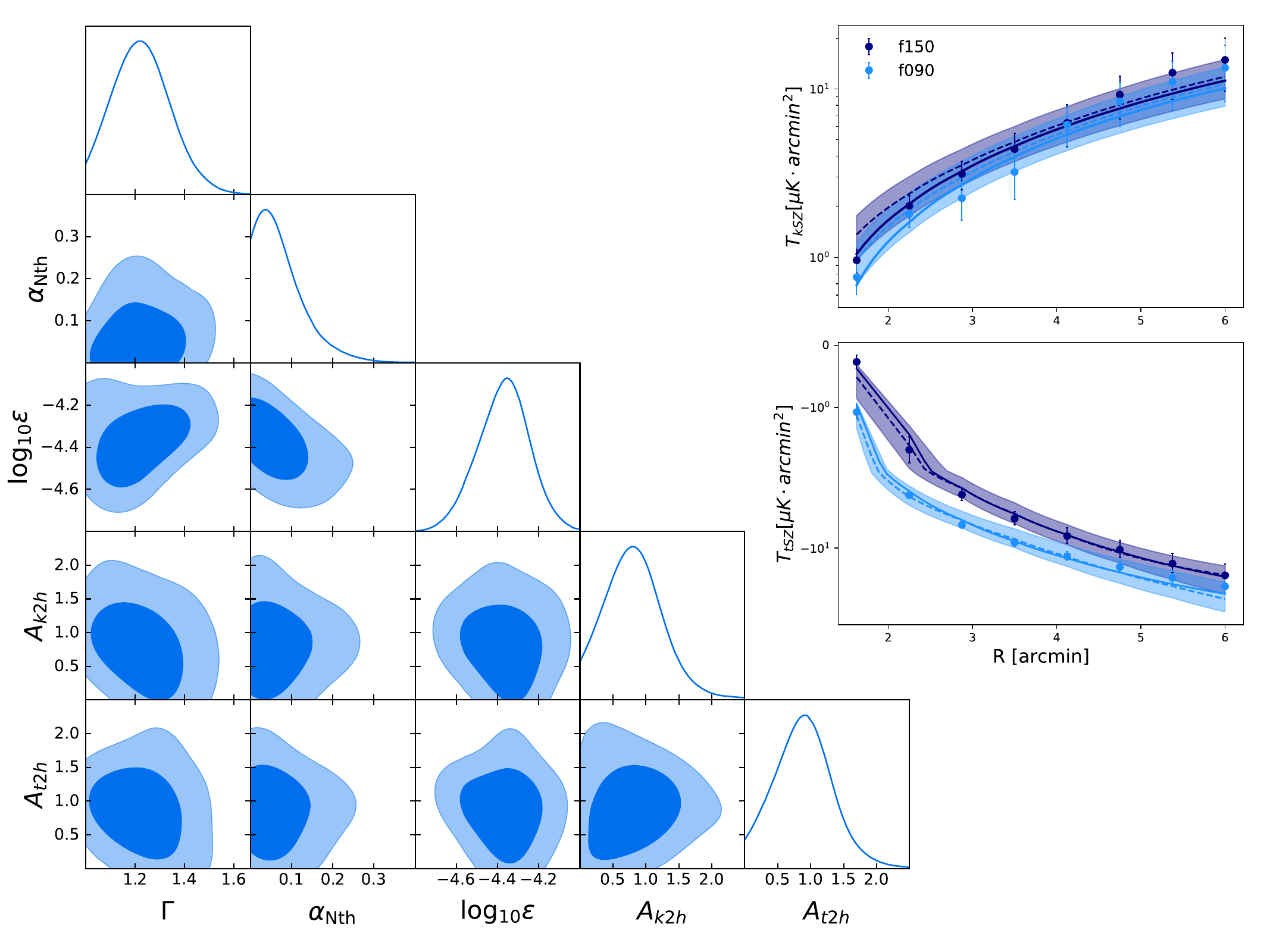}
  \caption{
  Constraints on the polytropic index, $\Gamma$,  the amplitude of the non-thermal pressure profile, $\alpha_{\rm Nth}$, the feedback efficiency parameter, $\epsilon$, and the amplitudes of the two-halo terms of the density and pressure profiles, $A_{\rm k2h}$  and $A_{\rm t2h}$, obtained by fitting the OBB model to combined kSZ and tSZ measurements by \cite{schaan+20}. The radial data have large correlations (see Fig. 7 in \cite{schaan+20}) that are accounted for in the analysis. The triangle plot shows one and two dimensional projections of the posterior probability distributions of the free parameters. \old{The dashed grey line shows a truncated Gaussian distribution as prior for $\Gamma$, centered at 1.2 and with sigma 0.2, in the range $1<\Gamma<5/3$.} We assume uniform priors within: $1<\Gamma<5/3$, $0<\alpha_{\rm Nth}<0.8$, $-4.8<\log_{\rm 10} \epsilon<-4.0$, $0<A_{\rm k2h}<5$, $0<A_{\rm t2h}<5$. The top right panels show the observed kSZ and tSZ  profiles (points) with the best-fit and the 2$\sigma$ range (2nd-98th percentiles) of the distribution of the models obtained from the MCMC chains. }
  \label{fig:OBBmcmc}
\end{figure*}

\begin{table*}
\centering 
\begin{tabular}{c c c c} 
\toprule 
Parameter & Description & Prior & Constraints (1$\sigma$) \\
\midrule
\multicolumn{4}{c}{OBB model} \\ 
\midrule
$\Gamma$ & Polytropic index &  [$1, 5/3$] & 
$1.2\pm0.1$  \\
\noalign{\vspace{1pt}}
$\alpha_{\rm Nth}$ & Non-thermal pressure norm. & [0.0, 0.8] & 
$0.04^{+0.07}_{-0.03}$  \\
\noalign{\vspace{1pt}}
$\epsilon$ & Feedback efficiency & $[10^{-4.8}, \,10^{-4.0}]$ & 
$(40\pm9)\times10^{-6}$  \\ 
\noalign{\vspace{1pt}}
$A_{\rm k2h}$ & Two-halo density amplitude &  [0,  5] &
$0.8\pm0.5$ \\ 
\noalign{\vspace{1pt}}
$A_{\rm t2h}$ & Two-halo pressure amplitude & [0, 5] &
$0.9^{+0.3}_{-0.5}$ \\
\bottomrule
\end{tabular}
\caption{
Marginalized constraints on the OBB parameters.   
} 
\label{tab:mcmc_OBB_par} 
\end{table*}

\begin{figure}
\centering
  \includegraphics[width=\columnwidth]{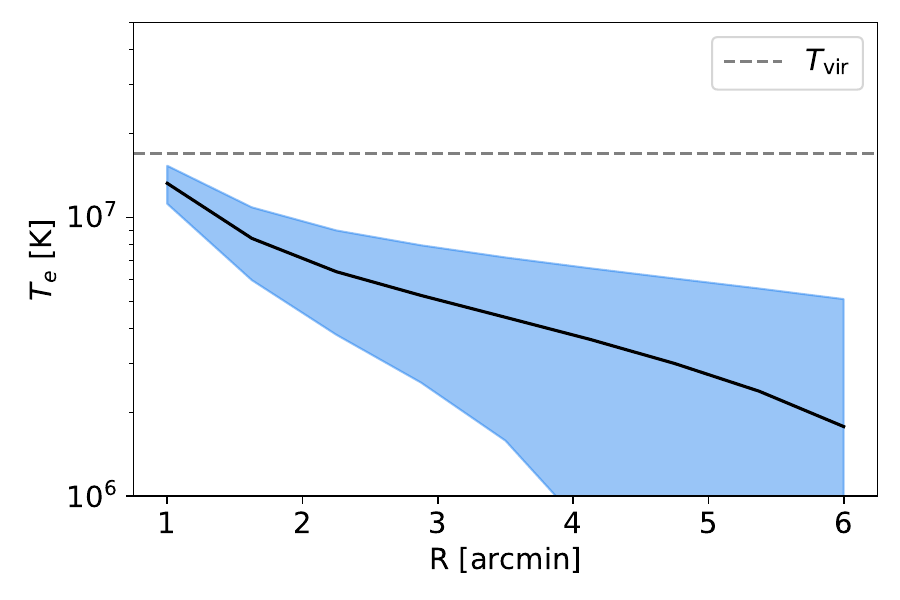}
  \caption{Average, inferred electron temperature profile of CMASS galaxies halos weighted by density obtained from the joint kSZ+tSZ fit to the OBB model using the MCMC chains. The black line is the median profile, the blue band indicates the 2$\sigma$ range of the models obtained from the MCMC chains. For comparison, the grey dashed line indicates the expected virial temperature for an isothermal sphere of mass equal to the the mean mass of our CMASS sample ($M_{\rm 200}=3.3\times10^{13}  M_\odot$), $T_{\rm vir}=1.7\times10^7$ K.
  The x-axis is converted to arcmins to ease the comparison to the density and pressure profiles. The average temperatures of the CMASS galaxies are closer to $10^{7}$K than they are to $10^{6}$K.}
  \label{fig:Te_prof}
\end{figure}

\subsection{GNFW Density}
\label{sec:gnfw_density}
\begin{figure*}
\centering
  \includegraphics[width=\textwidth]{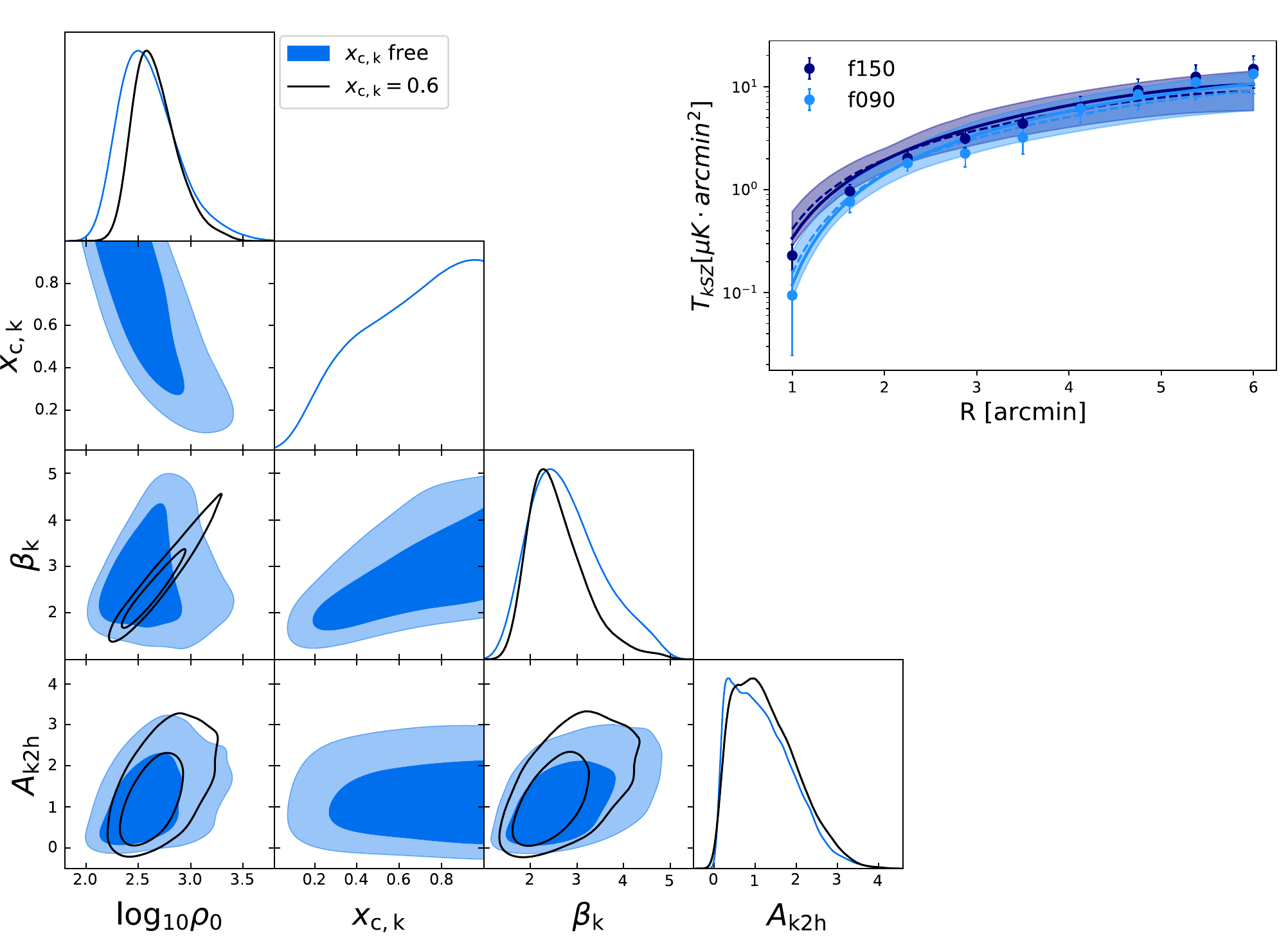}
  \caption{Constraints on the log-amplitude of the gas  density profile,  $\log_{\rm 10} \rho_0$, the core radius, $\xck$,the outer slope,  $\bk$, and the amplitude of the two-halo  term, $\aktwoh$, obtained by fitting the GNFW density model to kSZ measurements by \cite{schaan+20}. The radial measurements have large correlations (see Fig. 7 in \cite{schaan+20}) that that we take into account in our analysis. The corner plot shows one and two dimensional projections of the posterior probability distributions of the free parameters. We assume uniform priors on the parameters within: $1<\log_{\rm 10} \rho_0<5$, $0.1<\xck<1$, $1<\bk<5$, $0<\aktwoh<5$.
The top right panel shows the measured kSZ profile at f090 and f150 (circles) with the median (50th percentile, dashed curves) and the 2$\sigma$ (2nd-98th percentiles, bands) range of the models obtained from the MCMC chains. The solid lines indicate the best-fit model with $\chi^2$=20.2 and PTE=0.32. }
  \label{fig:GNFW_kSZ_mcmc}
\end{figure*}

Our GNFW density profile is defined by Eq.\ \ref{eq:rho_gnfw}. 
We fit for the density amplitude $\rho_0$, the core radius $\xck$, the power law index $\bk$ for the asymptotic fall-off of the profile, and the amplitude of the two-halo term $\aktwoh$. Given the considerable degeneracy of the parameters we fix two parameters that are sensitive to the profile properties at small radii, $\gk = - 0.2$, and $\alk = 1$, as in \cite{Battaglia+16}.
We define the likelihood combining the two density models for f090 and f150, accounting for the different beams and the correlated noise, in order to  jointly fit the kSZ measured in the two bands. 
We assume uniform priors on all the free parameters, in the ranges: $1<\log_{\rm 10} \rho_0<5$, $0.1<\xck<1.0$, $1<\bk<5$, $0<\aktwoh<5$. 
Table \ref{tab:gnfw_par} reports the marginalized constraints with $1\sigma$ error bars.
Figure \ref{fig:GNFW_kSZ_mcmc} shows the posterior contours of the GNFW density parameters. The top right panel shows the best-fit model (solid lines) and the $\pm2\sigma$ range of the models obtained from the MCMC over the kSZ data. The $\chi^2$ of the best-fit model (i.e. the minimum $\chi^2$) is 20.2, and the PTE is 0.32, indicating a good fit of the data. 
We next assess the significance of the detection of a two-halo term by our kSZ measurements. We find a best-fit amplitude of $\aktwoh=1.1_{-0.7}^{+0.8}$, indicating a $1.6 \sigma$ evidence, consistent with the values  obtained with the OBB fit.

The $\chi^2$ does not change if we reduce the number of free parameters by fixing the core radius to our best-fit value, $\xck=0.6$. In this case we get $\chi^2=20.1$, with three free parameters (PTE=0.33) and the same 2$\sigma$ range of the models obtained from the MCMC chains. The constraint on the log-amplitude improves from 13\%
to  7\% ($\log_{10} \rho_0=2.9\pm0.2$), while the constraints on $\bk$ and $\aktwoh$ do not substantially change.

We check the consistency between the kSZ radial profiles obtained  with the GNFW and the OBB models using a $\chi^2$ statistics. We find that they match within 1$\sigma$, with $\chi^2=15.8$ for 16 data points (PTE=0.47).

\subsection{GNFW Pressure}
\label{sec:gnfw_pressure}

\begin{figure*}
\centering
  \includegraphics[width=\textwidth]{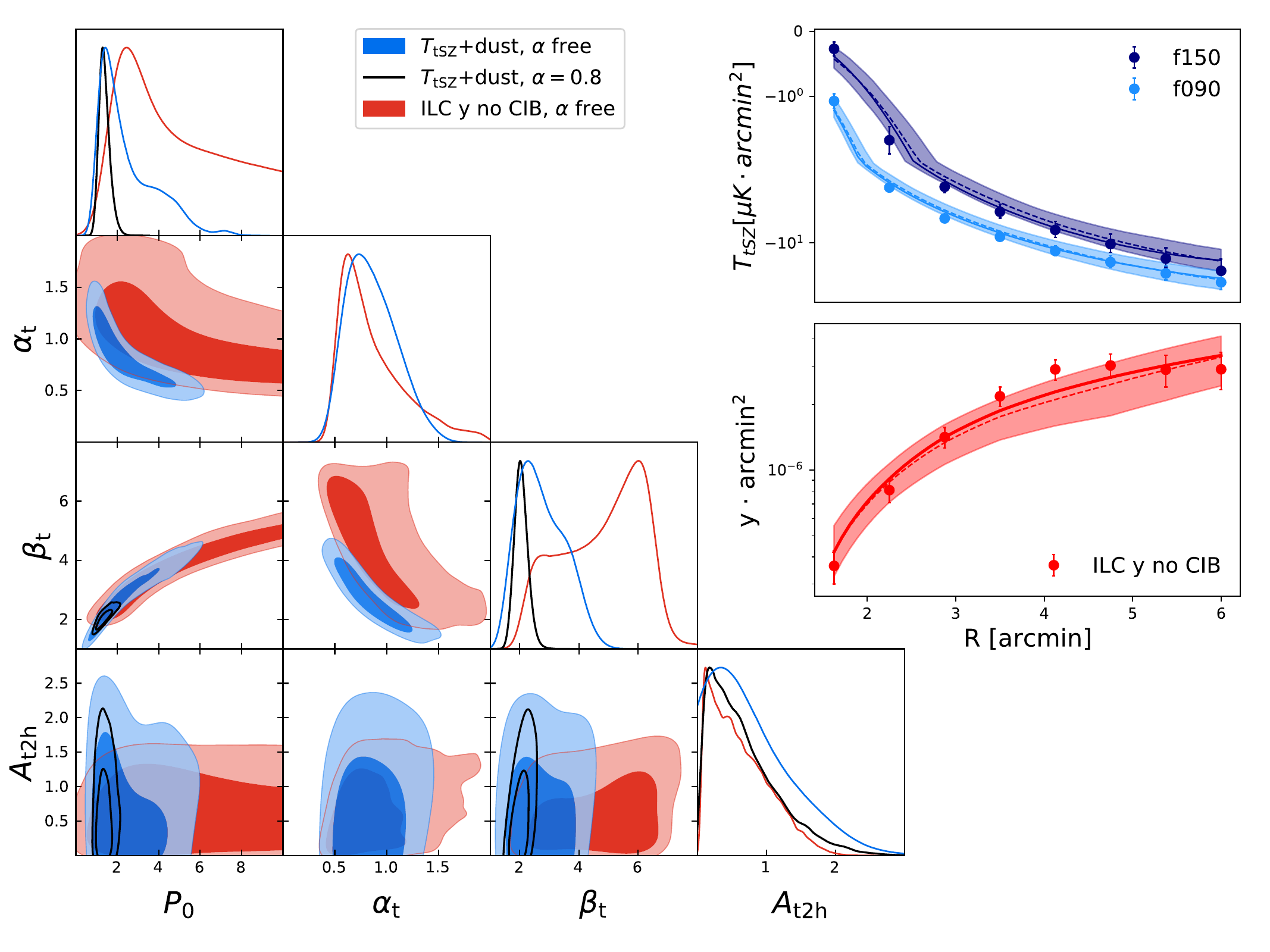}
  \caption{Constraints on the amplitude of the thermal  pressure profile, $P_0$, the intermediate slope, $\altt$, the outer  slope,  $\bt$, and  the amplitude of the two-halo term, $\attwoh$, obtained by fitting the GNFW pressure model to tSZ measurements by \cite{schaan+20}. The radial data have large correlations (see Fig. 7-8 in \cite{schaan+20}) that are accounted for in the analysis. The blue contours and lines show the fit of the GNFW thermal pressure+dust model to ACT and Herschel temperature measurements (see Fig.\ \ref{fig:dustfit} for the simultaneous constraints on the dust model), while in red is the fit of the GNFW thermal pressure model to Compton-y measurements obtained with CIB-deprojected maps .
The corner plot shows one and two dimensional projections of the posterior probability distributions of the free parameters. We assume uniform priors within: $0.1<P_0<30$, $0.1<\altt<2$, $1<\bt<10$, $0<\attwoh<5$. The top right panel shows best-fit (solid lines), the median (50th percentile, dashed lines) and the 2$\sigma$ (2nd-98th percentiles, bands) of the distribution of the models obtained from the MCMC chains.}
  \label{fig:GNFW_tSZ_mcmc}
\end{figure*}

Our GNFW pressure profile is defined by Eq.\ \ref{eq:P_gnfw}.  We fit for the amplitude $P_0$, the intermediate slope $\altt$, the power law index $\bt$ for the asymptotic fall-off of the profile, and the amplitude of the two-halo term $\attwoh$. There is significant degeneracy among all the GNFW parameters, so we fix two parameters sensitive to the profile properties at small radii, at the values suggested by previous cosmological simulations: $\gt = -0.3$ \cite{nagai+07, Battaglia+12b} and $\xct = 0.497 \times (M_{\rm 200}/10^{14} M_\odot)^{-0.00865} \times (1+z)^{0.731}$ \cite{Battaglia+12b}.
The tSZ measurements that we use to constrain our model include a contamination by the thermal emission from dust in our galaxy sample or in galaxies spatially correlated with it, which we take into account in our model. We fit the GNFW thermal pressure profile to the ACT data (f090, f150) and a dust modified black-body model to the ACT data and additional Herschel data from the H-ATLAS extragalactic survey \cite{eales+10} in the three bands centered at 600, 857, 1200 GHz, 
defining one combined likelihood. We present here constraints on the GNFW pressure profile and we refer to Appendix \ref{sec:dust} for details on the dust model.
We assume uniform priors on the parameters in the ranges: $0.1<P_0<30$, $0.1<\altt<2$, $1<\bt<10$, $0<\attwoh<5$. 
Table \ref{tab:gnfw_par} reports the marginalized constraints along with the $1\sigma$ errors. 
Figure \ref{fig:GNFW_tSZ_mcmc} shows in blue the posterior contours of the GNFW density parameters from the tSZ+dust fit. We do not find a noticeable correlation between the parameters of the GNFW and the parameters of dust model shown in Figure \ref{fig:dustfit}.
The top right panel shows the median ($\pm2\sigma$) range of the models obtained from the MCMC runs over the tSZ (+dust) data, and the best-fit corresponding to a minimum $\chi^2$ of 43.5 with PTE=0.45 (solid lines). In order to validate our model, and check that the constraints on the GNFW parameters are not determined by the dust fit, we also use measurements of the tSZ alone \cite{schaan+20} obtained with the internal linear combination (ILC) component-separated maps from \cite{Madhavacheril+19}, of the Compton-y with deprojected cosmic infrared background (CIB) from \emph{Planck} + ACT DR4. The result of fitting the CIB-deprojected Compton y maps is shown by the red contours in Figure \ref{fig:GNFW_tSZ_mcmc}. The best-fit to the Compton-y profile has a $\chi^2$ of 10.5 (PTE = 0.31) and matches within 1$\sigma$ the tSZ+dust fit.

A notable feature is the degeneracy between $\bt$ and $P_0$. This is a well known degeneracy and has been seen before in tSZ profile measurements \cite{Planck_V_2013}. We clearly observe this degeneracy in the fits to Compton-y, CIB-deprojected measurements and to a much lesser extent in the fits to tSZ+dust measurements. We attribute this difference to the fact that the measurement errors for the Compton-y CIB-deprojected case are larger, as a result of a slightly smaller area overlap with the CMASS sample. Moreover, the component separated maps  are not minimum variance, due to the nulling of the CIB, and they include ACT data up to 2015 only (DR4), as opposed to 2018 for our fiducial temperature maps (DR5).

The same GNFW form was previously used in \citep[e.g.,][]{Hill2013} to model the tSZ - CMB lensing cross-correlation.
There, the degeneracy was broken by keeping $\bt$ fixed, and the amplitude $P_0$ was that of the mean pressure profile of all halos in the Universe, weighted by their tSZ signal times their CMB lensing signal, instead of that of a specific galaxy sample like here.

The best-fit value $\attwoh=0.7_{-0.4}^{+0.8}$ indicates a preference for a non-zero two-halo term at $1.8 \sigma$ from tSZ measurements, consistent with the values obtained with the OBB fit. The measurement of the two-halo term alone is not new and previous studies have used stacked tSZ measurements to probe the distribution of hot gas in galaxy clusters and groups and to separate the one- and two-halo regimes.
Refs.~\cite{Vikram+17,hill+18} measured the two-halo term by analyzing the cross-correlation function between SDSS galaxy groups at a lower redshift ($z<0.2$) and \emph{Planck} y-maps.  They found evidence of both components in the most massive halos, $M\geq10^{13.5} h^{-1} M_\odot$, with a predominance of the two-halo term at $\gtrsim 2$~Mpc, and evidence of two-halo term alone for lower mass systems. 
Also using \emph{Planck} y-maps, the two-halo regime has now been constrained through the measurement of $\langle bP_e \rangle $, the halo bias-weighted mean electron pressure, with galaxy samples from the Dark Energy Survey \cite{Pandey2019}, a compilation of the 2MASS photometric redshift survey, WISE, and SuperCOSMOS \cite{Kouk2020}, and the DR14 SDSS release \cite{chiang+20a, chiang+20b}. Unlike previous work based on \emph{Planck}, the ACT data used here has a smaller beam, enabling us to study the pressure profiles in small group-sized halos, including both the one-halo and two-halo terms, at $z\sim0.6$. 

The goodness of the fit does not substantially change if we reduce the number of free parameters by fixing the intermediate slope to our best-fit value $\altt=0.8$. We  get in this case $\chi^2=40.1$ (PTE=0.60) and the same 2$\sigma$ distribution of the models obtained from the MCMC chains. The constraints on the amplitude get remarkably tighter, from 70\% to 20\% ($P_0=1.3^{+0.3}_{-0.2}$), and those on the outer slope improve from 33\%to 10\% ($\bt=2.0\pm0.2$), while the constraints on $\attwoh$ and on the parameters of the dust model remain essentially the same.

The tSZ radial profile that we obtain from fitting the GNFW model is consistent within 2$\sigma$ with the tSZ profile obtained for the OBB model. We get $\chi^2=27.9$ for 16 data points (PTE=0.03). This is a reasonable match considering that these are fits of different parametric models, each one having some degenerate parameters, and also taking into account our measurement errors. By neglecting the outermost measurements which have the largest error bars, we find a better match within 1.6$\sigma$, with $\chi^2=20.4$  for 14 data points (PTE=0.12).

\begin{table*}
\centering
\begin{tabular}{c c c c} 
\toprule 
Parameter & Description & Prior & Constraints (1$\sigma$) \\
\midrule
\multicolumn{4}{c}{GNFW density model}  \\ 
\midrule
$\log_{10}\rho_0$ & Log amplitude &  [1, 5] & \bm{$2.6_{-0.3}^{+0.4}$} \\ 
\noalign{\vspace{1pt}}
$\xck$ & Core radius & [0.1, 1] &  $0.6\pm0.3$  \\ 
\noalign{\vspace{1pt}}
$\bk$ & Outer slope & [1, 5] & $2.6_{-0.6}^{+1.0}$  \\ 
\noalign{\vspace{1pt}}
$\aktwoh$ & Two-halo term amplitude & [0, 5] & $1.1_{-0.7}^{+0.8}$ \\
\midrule
\multicolumn{4}{c}{GNFW pressure model} \\ 
\midrule
$P_0$  & Amplitude & [0.1, 30] & $2.0_{-0.8}^{+2.0}$  \\ 
\noalign{\vspace{1pt}}
$\altt$  & Intermediate slope & [0.1, 2] & $0.8_{-0.2}^{+0.3}$  \\ 
\noalign{\vspace{1pt}}
$\bt$  & Outer slope & [1, 10] & $2.6_{-0.7}^{+1.0}$ \\ 
\noalign{\vspace{1pt}}
$\attwoh$ & Two-halo term amplitude & [0, 5] & $0.7_{-0.4}^{+0.8}$ \\ 
\noalign{\vspace{1pt}}
\bottomrule
\end{tabular}
\caption{ Marginalized constraints on the GNFW parameters. } 
\label{tab:gnfw_par} 
\end{table*}

\section{Implications for optical weak-lensing observations}
\label{sec:lensing}

\begin{figure}[t]
\centering
  \includegraphics[scale=0.35]{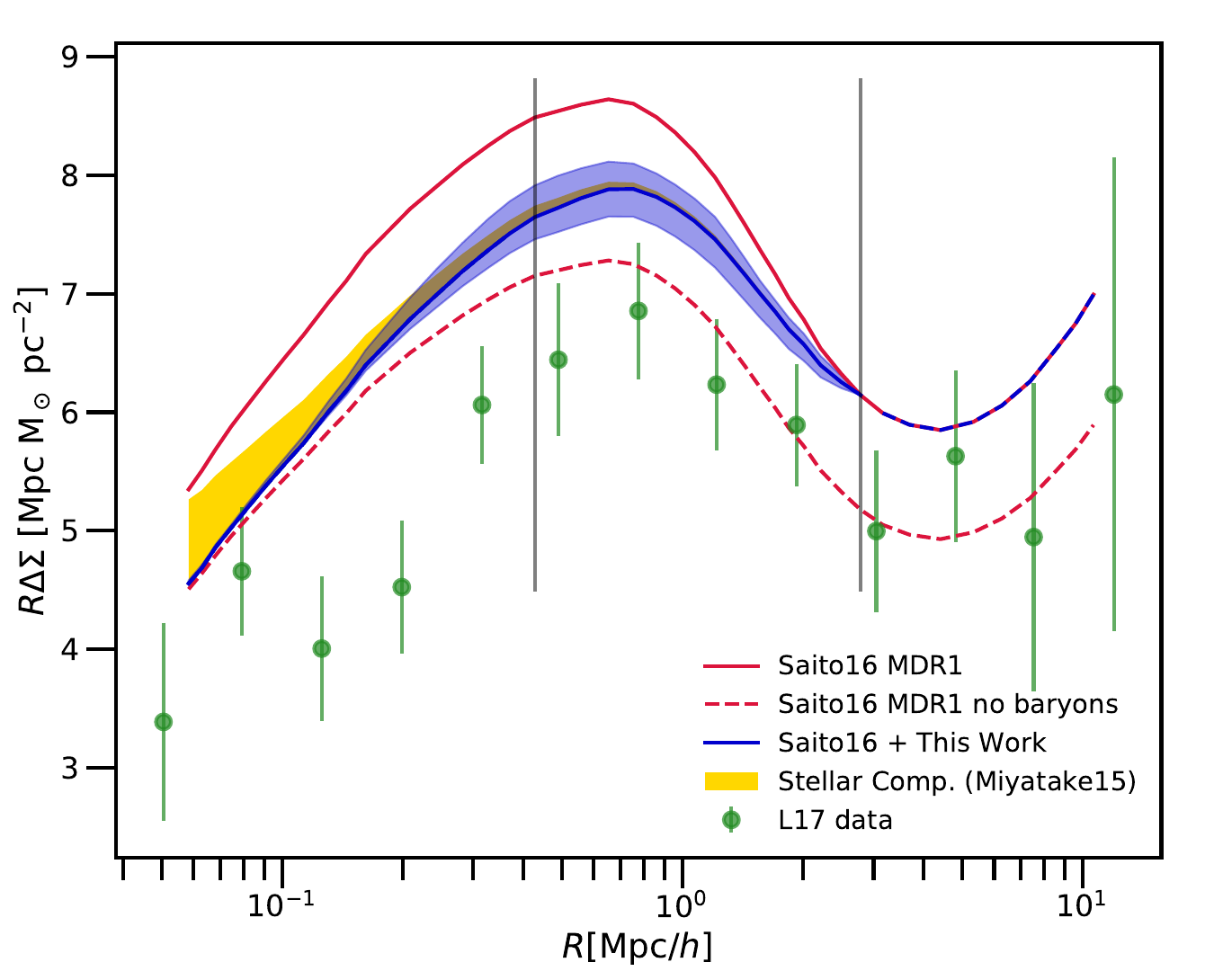}
  \caption{CMASS galaxy-galaxy lensing signal. Data from \cite{leauthaud2017} (green circles) are compared to HOD model predictions from \cite{saito+16} (MDR1, red line) and our model that include a baryons correction (blue line) to the MDR1. This correction uses the best fit density profile from kSZ measurements (Section \ref{sec:gnfw_density} and Figure~\ref{fig:GNFW_kSZ_mcmc}). The gold band illustrates the uncertainty in the model from the stellar component and the vertical grey lines show the radial range in which we have kSZ observations; outside this radial range we are extrapolating. The baryon correction that we estimated to the MDR1 model reduces the difference between the galaxy-galaxy measurements and HOD model predictions by half (50\%), but does not reconcile it. The dashed red line illustrates the maximum correction to the MDR1 model, which is to remove all baryons without altering the dark matter profile. This extreme model still does not reconcile this model and observations below 500 kpc$/h$.}
  \label{fig:lensing}
\end{figure}

The parametric GNFW model for the electron density profile we obtained from kSZ measurements serves as a first-order, empirical model for how baryons impact theoretical halo occupation distribution (HOD) models for optical weak-lensing measurements from the CMASS sample. Ref.~\cite{leauthaud2017} showed that their HOD model for the galaxy-galaxy lensing signal from CMASS over-estimated this signal compared to their measurements, concluding that ``lensing is low''. The details of their fiducial halo model (MDR1) are described in \cite{saito+16} and the parameters of their model are calibrated to provide the best fit to CMASS galaxy clustering measurements. 

Here, we do not attempt to disentangle the HOD from the individual profiles. Our best fit GNFW profile describes the ``HOD-convolved'' density profile. In other words our parametric GNFW model contains within it the underlying properties of the CMASS sample, like what fraction of the CMASS sample are central or satellite galaxies. Thus, it is indeed the relevant quantity for predicting the impact of baryons on galaxy weak lensing, since the weak lensing signal is also convolved with the same exact HOD.

With our HOD-convolved best fit we can straight-forwardly estimate the impact of baryons on the MDR1 model \cite{saito+16} by simply incorporating our parametric GNFW model for the electron density into it. 
The MDR1 model assumes that baryons trace the dark matter on all scales. We will use the MDR1 HOD model for the dark matter contribution to the galaxy-galaxy lensing measurement which uses a standard weak-lensing shear estimator, $\Delta \Sigma$. The projected mass density $\Sigma$ is related to $\Delta \Sigma$ through
\begin{equation}
    \Delta\Sigma(R) = \bar{\Sigma}(<R)- \Sigma(R),
    \label{eq:DS}
\end{equation}
where $\bar{\Sigma}(<R)$ is the mean projected mass density within projected radius $R$ and $\Sigma(R)$ is the surface mass density at $R$. We can split the total $\Delta\Sigma$ into a dark matter component ($\Delta\Sigma_{\rm DM}$ from MDR1) and baryon component ($\Delta\Sigma_{\rm b}$, obtained from our parametric GNFW model) such that $\Delta\Sigma_{\rm tot} = \Delta\Sigma_{\rm DM} + \Delta\Sigma_{\rm b}$. The $\Delta\Sigma_{\rm DM}$ is calculated by scaling the full $\Delta\Sigma$ from MDR1 by the dark matter fraction, $(\Omega_{\rm M} - \Omega_{\rm b})/\Omega_{\rm M}$. The $\Delta\Sigma_{\rm b}$ is calculated by projecting our best fit GNFW model for the electron density profile, 

\begin{equation}
    \Sigma_{\rm b}(R) \propto 2 \int_0^\infty \rho_{\rm gas}\left(\sqrt{R^2 + l^2}\right) {\rm d} l.
\end{equation}

Here $l$ is the line-of-sight direction on which we project, and the profile we fit is spherically symmetric so there is no preferred axis. The $\Delta\Sigma_{\rm b}(R)$ profile is calculated using Equation~\ref{eq:DS} once $\Sigma(R)$ is calculated. We normalize $\Delta\Sigma_{\rm b}(R)$ such that the baryon contribution to $\Delta\Sigma_{\rm tot}$ equals $f_b \Delta\Sigma_{\rm DM}$ at $R_{\rm max}$:

\begin{equation}
\Delta\Sigma_{\rm b}(R)
\rightarrow
\Delta\Sigma_{\rm b}(R)
\times
\frac{f_b \Delta\Sigma_{\rm DM}(R_{\rm max})}{\Delta\Sigma_{\rm b}(R_{\rm max})}.
\end{equation}
Here $R_{\rm max}$ is the maximum angular radial bin for which we have a kSZ measurement.
To summarize, we assumed that all the baryons are present within the maximum radius that we measured and beyond this radius the baryons trace the dark matter. We note that this model does not include the effect of the dark matter profile rearranging itself in response to the new baryon profile, often referred to as a ``back-reaction'' to the baryons (e.g.  \cite{vD2011,tng2018}). We expect this to be a second-order correction to the model (supported by simulations e.g. \cite{tng2018}), smaller than the baryonic effect we included.

Figure \ref{fig:lensing} shows the original galaxy-galaxy lensing measurement from \cite{leauthaud2017} with green points and error bars, along with the original MDR1 HOD model from \cite{saito+16} shown as a red line. Our new estimate for the MDR1 halo model with a baryon correction coming from our kSZ profile measurements is shown in blue and the corresponding blue band illustrates the $2\sigma$ uncertainty obtained by sampling the best fit GNFW MCMC chains. The dashed red line illustrates what the \cite{saito+16} HOD model would predict if one were to remove all the baryons. 
This ``no-baryons'' curve sets a lower limit to the MDR1 HOD model of the galaxy-galaxy lensing signal, in the absence of a modification to the dark matter profile.
The yellow band shows the $2\sigma$ upper limit from the stellar component of $\Delta\Sigma_{\rm tot}$ following the calculations from \cite{Miyatake2015} and the vertical grey lines show the radial range of kSZ measurements from \cite{schaan+20}. Our estimates for the inner radii beyond the grey boundary are extrapolations of the model. At these radii the uncertainty from the stellar component is dominant.

Our empirical model for the baryon correction to the MDR1 halo model does reduce the difference between the galaxy-galaxy lensing measurement of the CMASS sample \cite{leauthaud2017} and the predicted signal from the \cite{saito+16} MDR1 HOD model, which is calibrated to the clustering of the CMASS sample. At its largest our baryon correction accounts for half the difference (50\%). However, the lensing measurements still fall below our model on all scales. Even assuming an extreme baryon correction model where all the baryons are removed from MDR1 HOD model, without altering the dark matter profile, the measured lensing signal is still below the model on scales of 500 kpc$/h$ and less. The impact of baryons is one of many effects considered in \cite{leauthaud2017}, the others being measurement systematics, sample selection, assembly bias, and extensions to our concordance cosmological model. It is likely that a combination of these effects is responsible for the low lensing signal (e.g. \cite{Lange2019}), since baryonic effects cannot explain the entire difference.

\section{Comparison to simulations}
\label{sec:sims}

Our measured kSZ and tSZ profiles from ACT$+$CMASS \cite{schaan+20} offer a new opportunity to test current cosmological simulations \cite{Battaglia+17,BH2019,DSR} and the sub-grid physics models they include to capture physical processes like feedback from stellar sources and AGN. Since these measurements are new, current simulations are not calibrated to match them, and thus the simulations permit a genuine prediction for these tSZ and kSZ CGM profiles. 

\begin{figure*}
\centering
  \includegraphics[scale=0.55]{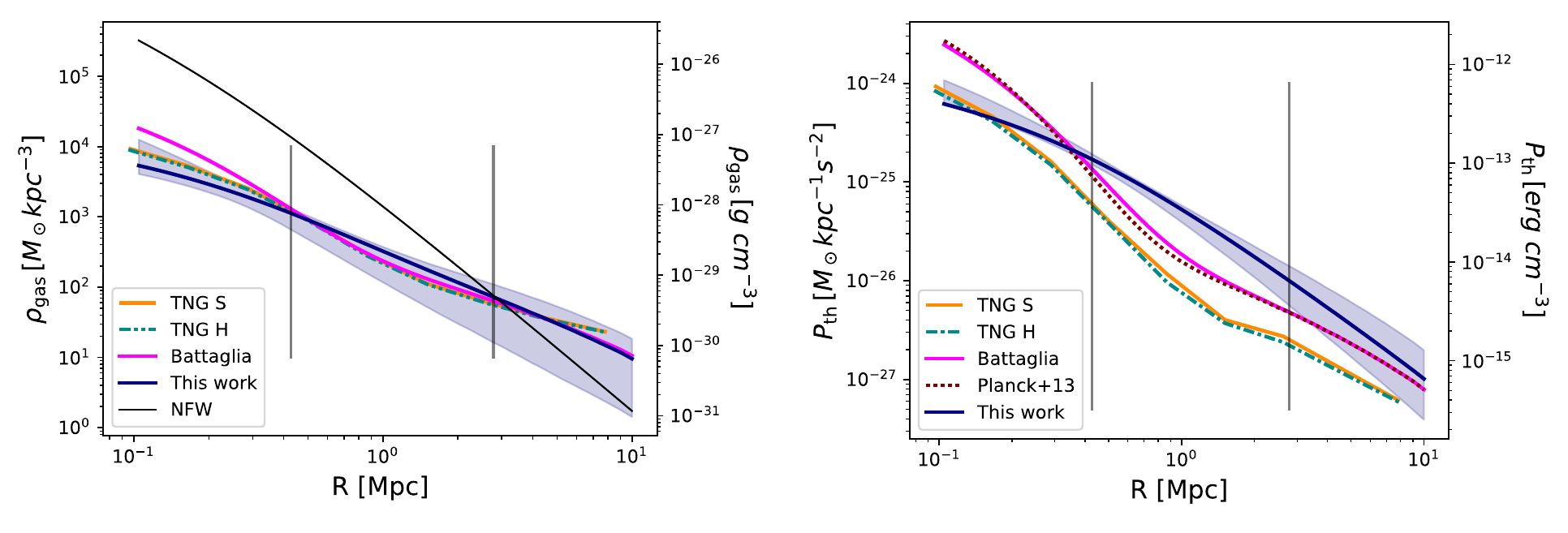}
  \includegraphics[scale=0.6]{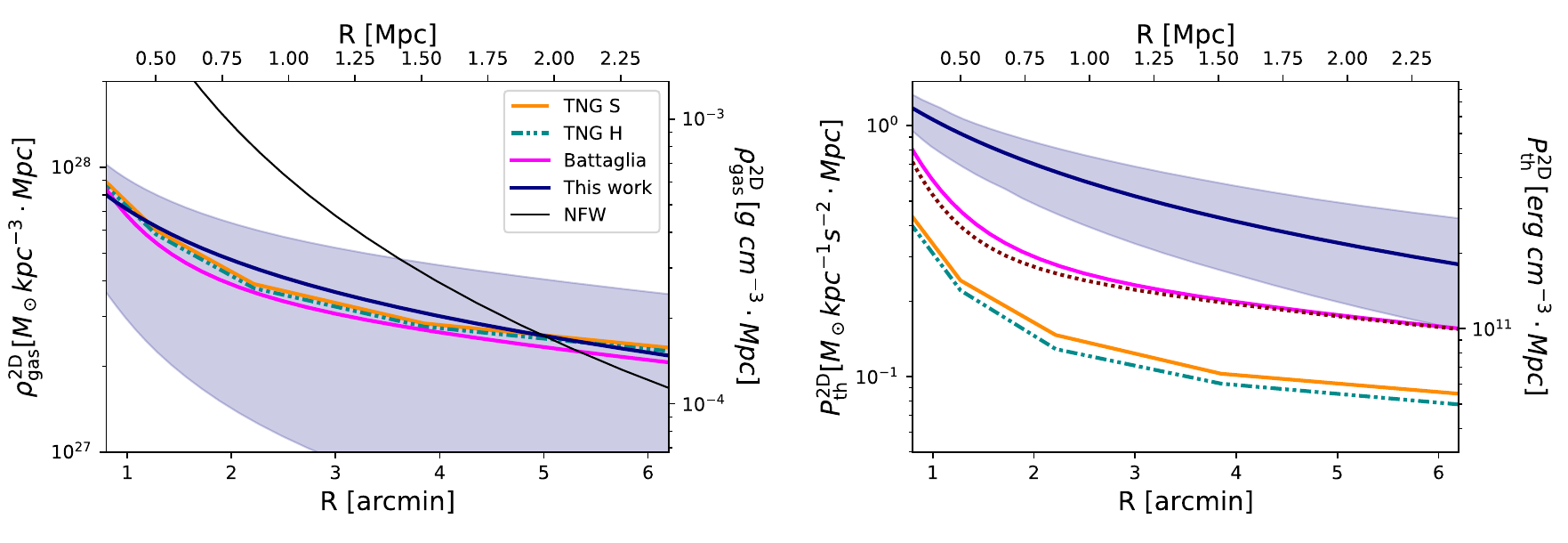}
  \includegraphics[scale=0.56]{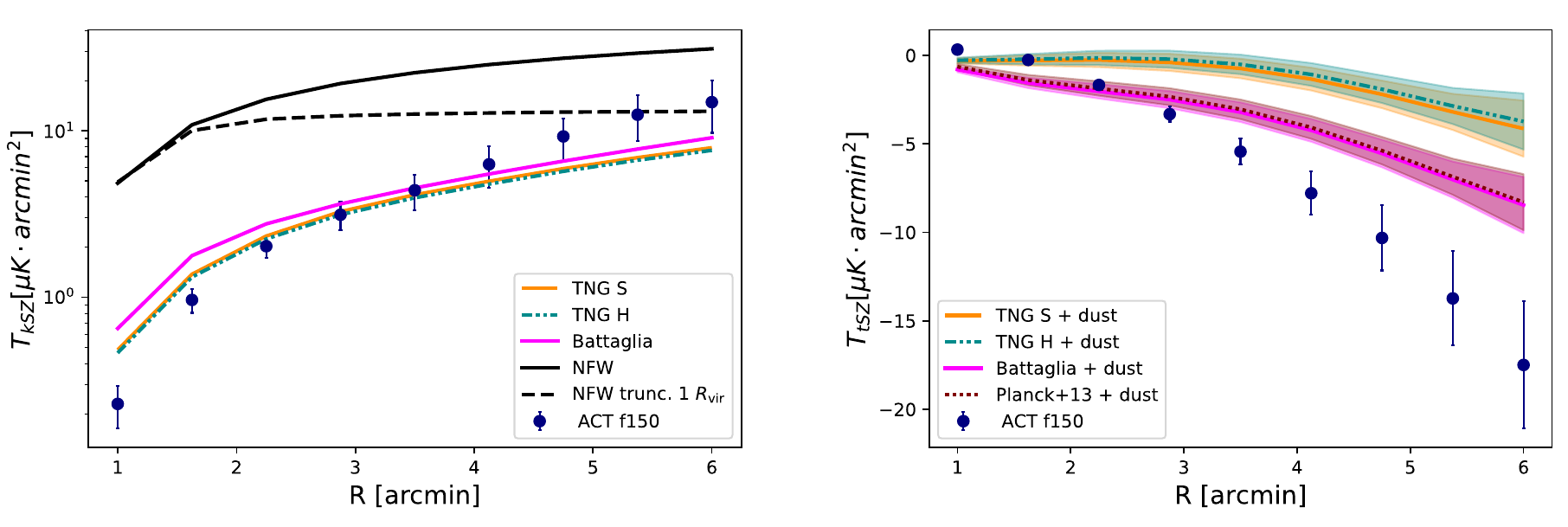}
  \caption{
  Top: comparison of our best-fit gas density (left) and thermal pressure (right) profiles (blue curves and $2\sigma$ bands) with the related profiles from two cosmological simulations: \cite{BBPSS2010} (magenta) and Illustris/TNG \cite{tng2018} (orange and green), and a NFW profile \cite{nfw} (black). 
  For the pressure profile, we also show the Planck 2013 \cite{Planck_V_2013} best fit (maroon dotted line).
  We show average profiles, where each halo contribution is weighted by its mass according to the mass probability density function of the CMASS catalog used in this work, and at the same redshift ($z=0.55$). We select red galaxies from TNG and show both stellar mass- (orange) and halo mass-weighted average profiles.
  The vertical grey lines enclose the range where  we measure the kSZ and the tSZ.
  Middle: projected density and pressure profiles, for comparison purposes.
  Bottom: comparison of the profiles projected into the kSZ (left) and tSZ (right) observable space with the measurements by \cite{schaan+20} in the ACT f150 band (blue points and 1$\sigma$ error bars). The projections of the simulated and the NFW profiles account for the convolution with the ACT beam and the aperture photometry filtering, as described in Section \ref{sec:modelobs}. The black dashed curve shows the NFW profile truncated at the virial radius. The tSZ simulated profiles also include the dust correction from our ACT+Herschel measurements (2$\sigma$).  
}
  \label{fig:sims}
\end{figure*}

We use predicted density and pressure profiles from Illustris TNG \cite{tng2018} and the \cite{BBPSS2010} simulations, and a NFW density profile \cite{nfw}, shown in the top panel of Figure~\ref{fig:sims}. For the TNG simulations, we use the simulation snapshot data that matches the mean redshift of the CMASS sample most closely. We further model the CMASS sample by selecting halos from Illustris TNG that were ``red'' in color, according to Illustris TNG, and we weight each halo's contribution by its mass, for both the stellar and halo mass distribution to match the observed sample's stellar (TNG S) and halo mass (TNG H) distributions, respectively. These two halo selections are meant to capture the uncertainty in the stellar mass to halo mass relation used for the CMASS sample and they are a decent metric for the uncertainty in the modeling of the CMASS sample with TNG. Red galaxies within Illustris TNG were selected to have colors $\rm sdss\_g-\rm sdss\_r \geq 0.6$ \cite{nelson+18}. For the \cite{BBPSS2010} simulations we use the fitting formulas from \cite{Battaglia+12b} and \cite{Battaglia+16}, include the mean redshift of the CMASS sample and weight the mass dependence according to the halo mass distribution of CMASS. These fitting formulas are extrapolated to lower masses, since \cite{BBPSS2010} do not resolve halos down to the masses of the CMASS sample. 
We also show the best fit GNFW pressure profile from {\it Planck} \cite{Planck_V_2013}, extrapolated to the CMASS masses, which is very similar to the predictions of \cite{BBPSS2010}. 
The middle panel of Figure~\ref{fig:sims} shows the projected density and pressure profiles for comparison purposes.
In the bottom panel of Figure~\ref{fig:sims} we compare the measured kSZ and tSZ profiles in the f150 frequency band from ACT+CMASS \cite{schaan+20} to the simulation predictions, obtained by convolving the projected profiles with the ACT beam and applying the aperture photometry filter as described in Sec.\ \ref{sec:modelobs}.
See \cite{schaan+20} for a discussion about the uncertainty in the NFW modeling.
The tSZ simulated profiles also include the dust correction from our ACT+Herschel measurements (the solid curves are given by the sum of the tSZ simulated profiles and our best-fit dust model, the bands enclose the 2$\sigma$ range).  
For the density-to-kSZ projection (Eqs.\ \ref{eq:ksz}-\ref{eq:el_prof}) we use for the simulated profiles the same $v_r$ that we use for our models, $v_r=1.06\times10^{-3} c$, which is the value computed in the linear approximation at $z=0.55$.
We do not include uncertainties on the profile predictions from simulations. These predictions are weighted averages over hundreds of simulated halos, thus the errors on these averages scales like $1/\sqrt{N}$, where $N$ is the number of halos and they are dwarfed by the measurement error.

At the smaller radii the simulations do a decent job of matching the signal, although the pressure profiles over-predict the signal there (higher negative values of $T_{\rm tSZ}$ correspond to higher values of the thermal energy/pressure).
These first four angular radial bins below 3 arcmin are within two virial radii (defined as an average overdensity of 200 with respect to the mean matter density) and have been probed indirectly in the past, through integrated gas density and Compton-y measurements. Given that some of these observational constraints were not available prior to calibrating these simulations it is not surprising that there are large differences between the simulations and the \cite{schaan+20} observations.
\new{At larger radii the pressure profiles predicted by both simulations are significantly lower than the measurements.} 
The \cite{BBPSS2010} simulations have the lowest $\chi^2$ of 12.7 (PTE=0.03), while the $\chi^2$s for the TNG S and TNG H predictions are 20.2 (PTE=0.00) and 23.1 (PTE=0.00) respectively. \new{For the density profile the total $\chi^2$ values for Illustris TNG and the \cite{BBPSS2010} simulations are very similar and range from 8.2-10.1, with PTEs 0.07-0.15.}
Unlike the smaller radii, this radial range is completely unexplored. Clearly the simulations are under-predicting the gas temperature at these radii. This suggests that the sub-grid stellar and AGN feedback models these simulations use to stop over-cooling in the center and remove low entropy gas does not sufficiently heat the gas in the outer regions of the CGM. 

There are numerous reasons why simulations could under predict the amount of CGM pressure at these larger radii. For example, the numerical methods chosen to inject energy and how that energy is allowed to propagate through the CGM will impact the thermodynamic properties of the CGM on all scales. Previous comparisons between simulations and tSZ profiles have been mostly limited to higher masses, like galaxy clusters \citep[e.g.,][]{Planck_V_2013,McDonald2014,Lebrun2014,Barnes2017}, where the simulations matched the observations across a large range in radii. This implies that the current sub-grid models for energy injection and numerical method for propagation are sufficient at describing the ICM but do not sufficiently heat the CGM at radii beyond the virial radius.

Predicting the gas profiles on these large scales in $10^{13} M_\odot$ galaxies is challenging as they have smaller potential wells than galaxy clusters and their CGM is more susceptible to small changes in the feedback modeling. A challenge going forward for cosmological simulations will be to sufficiently heat the CGM at radii beyond the virial radius without completely unbinding the CGM from halos at this mass scale. We look forward to investigating additional data from other current cosmological simulations and potentially enabling further refinement of the current sub-grid feedback models. 

\section{Summary and conclusions}
\label{sec:summary}

We present constraints on the gas thermodynamics of CMASS galaxies using kSZ and tSZ cross correlation measurements from \cite{schaan+20}.

Combining kSZ and tSZ measurements we constrain the efficiency of feedback, in terms of thermal energy injected into the gas from AGN and supernovae, $\epsilon=(40\pm9)\times10^{-6}$ (1$\sigma$), which we robustly estimate with a 23\% relative uncertainty. From this result we estimate that the energy injected by feedback is 30\% of the total binding energy of the system, which is consistent with the 25\% value calculated from \cite{BBPSS2010} simulations.
We find an upper limit for the amplitude of the non-thermal pressure profile,
$\alpha_{\rm Nth}<0.2\ (2\sigma)$, indicating that less than 20\% of the total electron pressure within $R_{\rm 200}$ is due to a non-thermal component.

Thanks to the high significance data and the small ACT beam, we are also able to study the gas density and pressure profiles in the group-sized CMASS halos at $z\sim0.6$ in both the one-halo and two-halo regime.
We use the kSZ measurements to constrain the amplitude and shape of a generalized NFW model of the gas density profile, 
finding best fit parameters: 
\newm{$\mathrm{log_{10}} \rho_0=2.6_{-0.3}^{+0.4}$}, 
$\xck=0.6\pm0.3$, 
$\bk=2.6_{-0.6}^{+1.0}$, 
$\aktwoh=1.1_{-0.7}^{+0.8}$.
From the tSZ(+dust) measurements we constrain a generalized NFW model of the thermal pressure profile,
with best fit parameters
$P_0=2.0_{-0.8}^{+2.0}$,
$\altt=0.8_{-0.2}^{+0.3}$,
$\bt=2.6_{-0.7}^{+1.0}$,
{$\attwoh=0.7_{-0.4}^{+0.8}$} 
(1$\sigma$ error bars).

Using our best fit density profile from the kSZ measurements we estimate the baryon correction to the \cite{saito+16} HOD model of the CMASS galaxy-galaxy lensing signal, which is calibrated to match the CMASS clustering measurements. We find that including our baryon correction reduces but does not fully reconcile the difference with this galaxy-galaxy lensing measurement \cite{leauthaud2017}.

We also use the kSZ and tSZ measurements to directly test cosmological simulations with sub-grid physics modelling that are clearly not calibrated to match our new SZ observations of the CGM. The predicted density and pressure profiles from Illustris TNG \cite{tng2018} and the \cite{BBPSS2010} simulations match our data decently at $\lesssim 2 R_{\rm vir}$\new{,  while at larger radii the simulations underestimate the pressure.} 
 We interpret these underestimates of the CGM pressure, seen in a range so far unexplored, as inadequacies of the sub-grid stellar and AGN feedback models. We will continue to investigate additional data from other current cosmological simulations, which will potentially enable future simulations to refine their sub-grid feedback models. 

These combined kSZ and tSZ profile measurements have ushered in a new era of modeling and inference, especially thanks to the improvement from the past $\sim 4\sigma$ to the current $8\sigma$ kSZ  measurements. However, the interpretation of these higher signal-to-noise measurements require attention to both measurement and modeling systematics.
The main measurement systematic in the interpretation of the tSZ signal is the thermal dust emission from CMASS galaxies. Therefore we include in our thermal pressure model a model for the dust contribution that we constrain by stacking on ACT and Herschel data.

In order to optimally describe the observations, we forward model our theoretical density and pressure profiles to our observations. This includes convolving our models with the map beam profiles and applying the tSZ band-pass responses, which are computed for each frequency band. For the theoretical modeling we include both the one-halo and two-halo contributions and mass-weight our density and pressure profiles.
With upcoming higher signal-to-noise data, precisely modeling the HOD and the selection of the galaxy sample will be crucial to meaningfully compare measurements and hydrodynamical simulations.

This work demonstrates the power of joint tSZ and kSZ cross-correlation measurements in studying the distribution of baryons in the CGM of CMASS galaxy groups, especially in low-density environments and out to the outskirts, where they can reveal information about assembly history and the feedback processes.
Future CMB observations such as the Simons Observatory \cite{SO}, CCAT-Prime \cite{ccatp}, CMB-S4 \cite{S4} and spectroscopic surveys of the large-scale structure like the Dark Energy Spectroscopic Instrument (DESI \cite{desi}), the Subaru Prime Focus Spectrograph (PFS \cite{pfs}) and Euclid \cite{euclid}, will improve the precision in this radial range even more, with higher sensitivity, larger sky and frequency coverage, and larger galaxy samples, enabling more detailed studies across multiple sub-samples of mass, redshift, and galaxy properties. 

\acknowledgments{
The authors thank the anonymous referee for their helpful and constructive comments which improved the paper.
This work was supported by the U.S. National Science Foundation through awards AST-1440226, AST0965625 and AST-0408698 for the ACT project, as well as awards PHY-1214379 and PHY-0855887. Funding was also provided by Princeton University, the University of Pennsylvania, and a Canada Foundation for Innovation (CFI) award to UBC. ACT operates in the Parque Astron\'{o}mico Atacama in northern Chile under the auspices of the Comisi\'{o}n Nacional de Investigaci\'{o}n Cient\'{i}fica y Tecnol\'{o}gica de Chile (CONICYT).
The Flatiron Institute is funded by the Simons Foundation. NB acknowledges support from NSF grant AST-1910021. NB and JCH acknowledge support from the Research and Technology Development fund at the Jet Propulsion Laboratory through the project entitled ``Mapping the Baryonic Majority''. E.S. is supported by the Chamberlain fellowship at Lawrence Berkeley National Laboratory. S.F. is supported by the Physics Division of Lawrence Berkeley National Laboratory.
EC acknowledges support from the STFC Ernest Rutherford Fellowship ST/M004856/2 and STFC Consolidated Grant ST/S00033X/1, and from the Horizon 2020 ERC Starting Grant (Grant agreement No 849169).
RD thanks CONICYT for grant BASAL CATA AFB-170002. 
DH, AM, and NS acknowledge support from NSF grant numbers AST-1513618 and AST-1907657. 
MH acknowledges support from the National Research Foundation of South Africa.
JPH acknowledges funding for SZ cluster studies from NSF AAG number AST-1615657.
KM acknowledges support from the National Research Foundation of South Africa. 
CS acknowledges support from the Agencia Nacional de Investigaci\'on y Desarrollo (ANID) through FONDECYT Iniciaci\'on grant no.~11191125.
}

\bibliographystyle{prsty.bst}
\bibliography{references.bib}

\appendix
\section{Two-halo term}
\label{sec:twohalo}
We investigate the contribution to the halo gas profiles from neighboring halos, known as ``two-halo term''. We are interested in the two-halo kSZ signal observed from our cross-correlation analyses. In order to estimate this contribution, we construct an analytical model of the signal following the halo model of \cite{Vikram+17} (based on the formalism of \cite{CooraySheth2002}). 
The total halo-density correlation function describes the average excess density around halos with respect to random locations in the Universe, as a function of the comoving distance from the halo centre ($r$). This has both a one-halo and a two-halo contribution:
\begin{equation}
    \xi_{h,\rho} (r|M) = \xi_{h,\rho}^{\rm one\mhyphen halo} (r|M) + \xi_{h,\rho}^{\rm two\mhyphen halo} (r|M) \,,
\end{equation}
where $\xi_{h,\rho}^{\rm one\mhyphen halo} (r|M) = \rho_{\rm gas} (r|M)$ is the gas density profile of the halo itself (or equivalently the halo pressure profile $P_{\rm th} (r|M)$), while $\xi_{h,\rho}^{\rm two\mhyphen halo} (r|M)$ is the contribution from correlated neighboring halos.

In order to calculate the two-halo term, the first step is to compute the Fourier transform of the density profile around a neighboring halo:
\begin{equation}
    u_{\rho}(k,M) = \int_0^\infty dr~4\pi r^2 ~\frac{\sin(kr)}{kr} ~\rho_{\rm gas} (r|M) \,,
\end{equation}
assuming a spherically symmetric density profile. We then compute the two-halo contribution to the halo–density power spectrum:
\begin{equation}
    P_{h,\rho}(k) = b(M) P_{\rm lin}(k) \int_0^\infty dM' ~ \frac{dn}{dM'} ~b(M') ~u_\rho(k,M') \,.
\end{equation}
Here, $M$ is the mass of the halo of interest, the integral is over the masses $M'$ of the neighbour halos, in the range $10^{10}-10^{15} M_\odot$. $P_{\rm lin}(k)$ is the linear density power spectrum computed with the fit of \cite{EisensteinHu1998}, $dn/dM$ is the mass function of the neighboring halos that we compute from \cite{ShethTormen2002}, $b(M)$ is the linear bias factor of the halo of mass $M$, from \cite{ShethMoTormen2001}. 

Finally, we Fourier transform the weighted power spectrum to get the two-halo term of the correlation function:
\begin{equation}
    \xi^{\rm two-halo}(r|M) = \int_0^\infty \frac{dk}{2\pi^2} ~k^2 ~\frac{\sin(kr)}{kr} ~W(k) ~P_{h,\rho}(k)\,.
\end{equation}
We consider that for the kSZ signal, the two-halo contribution comes from halos within the correlation length of the linear velocity field, $r_{\rm corr}$, which is approximately 50 Mpc. Therefore we apply a window function $W(k) = 1$  for $k>1/r_{\rm corr}$, 0 elsewhere. 

\begin{figure*}
\centering
  \includegraphics[scale=0.5]{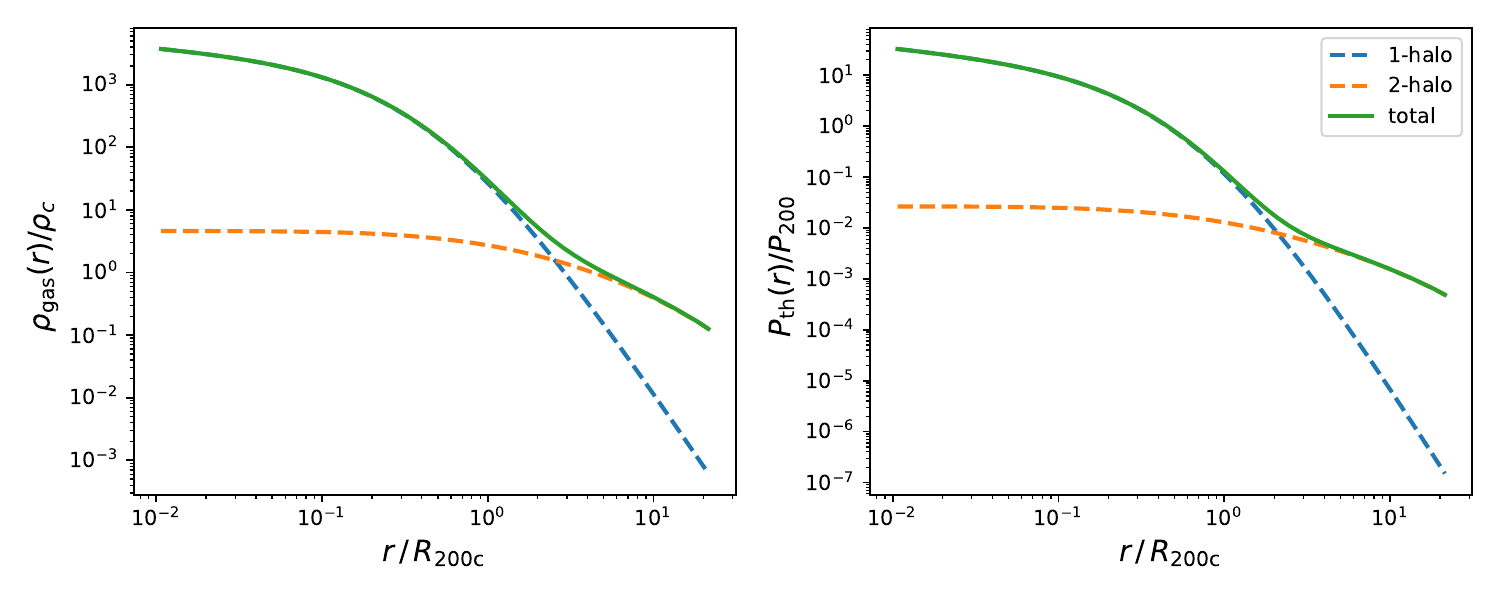}
  \caption{One and two halo terms contributing to the density (left) and pressure (right) profiles of a halo of $3\times 10^{13}M_\odot$ at $z=0.55$. The contribution of the two-halo term is not negligible above  $\sim 2 R_{200c}$.}
  \label{fig:2halo}
\end{figure*}

Figure \ref{fig:2halo} shows the fiducial two-halo term profiles that we calculate  for the gas density and thermal pressure of an average CMASS halo. For our fits, we include a free parameter in front of those terms that scales the amplitude, so that our final models of density and pressure profiles are:
\begin{equation}
\begin{aligned}
 \rho(r) = \rho_{\rm one\mhyphen halo}(r) + \aktwoh \,\rho_{\rm two\mhyphen halo}(r) \,, \\
 P(r) = P_{\rm one\mhyphen halo}(r) + \attwoh \, P_{\rm two\mhyphen halo}(r) \,,
\end{aligned}
\end{equation}

where $\aktwoh$ and $\attwoh$ are the free amplitude parameters for the density and pressure two-halo terms, respectively.

\section{Dust emission}
\label{sec:dust}
\begin{figure}
\centering
  \includegraphics[scale=0.28]{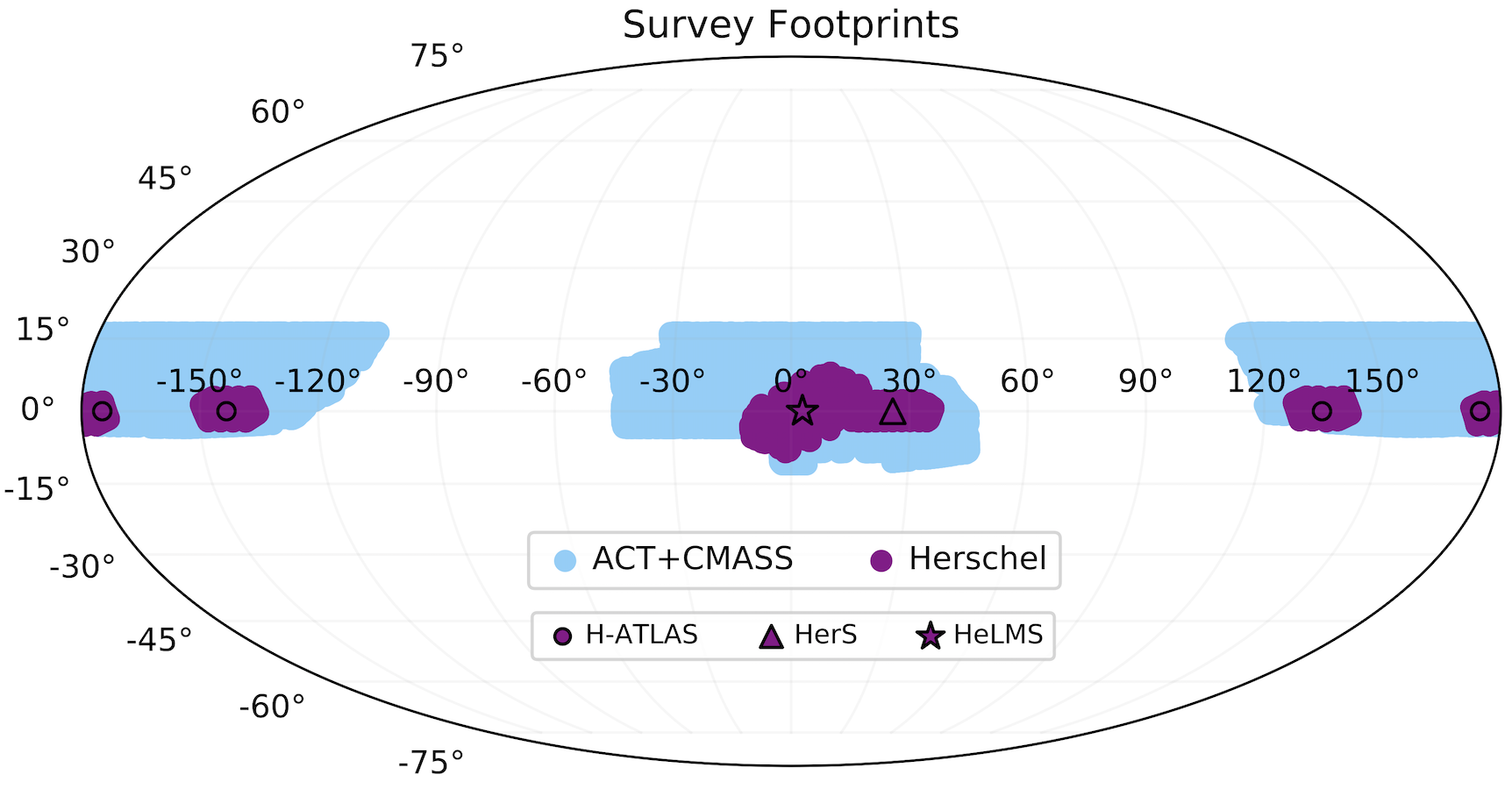} 
  \caption{Survey footprints, in equatorial coordinates, for the ACT+CMASS (blue) and overlapping Herschel (magenta) regions: three H-ATLAS/GAMA fields (circles), HerS (triangle) and HeLMS (star). We use the H-ATLAS/GAMA data only to estimate the dust emission (see text).}
  \label{fig:footprint}
\end{figure}

\begin{figure*}
\centering
  \includegraphics[scale=0.3]{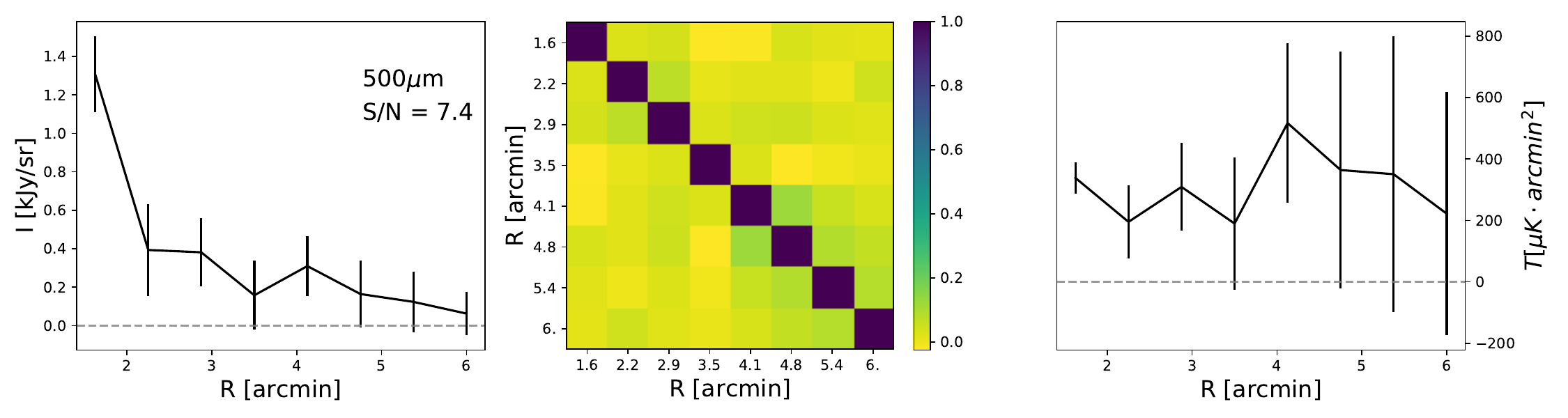} 
  \includegraphics[scale=0.3]{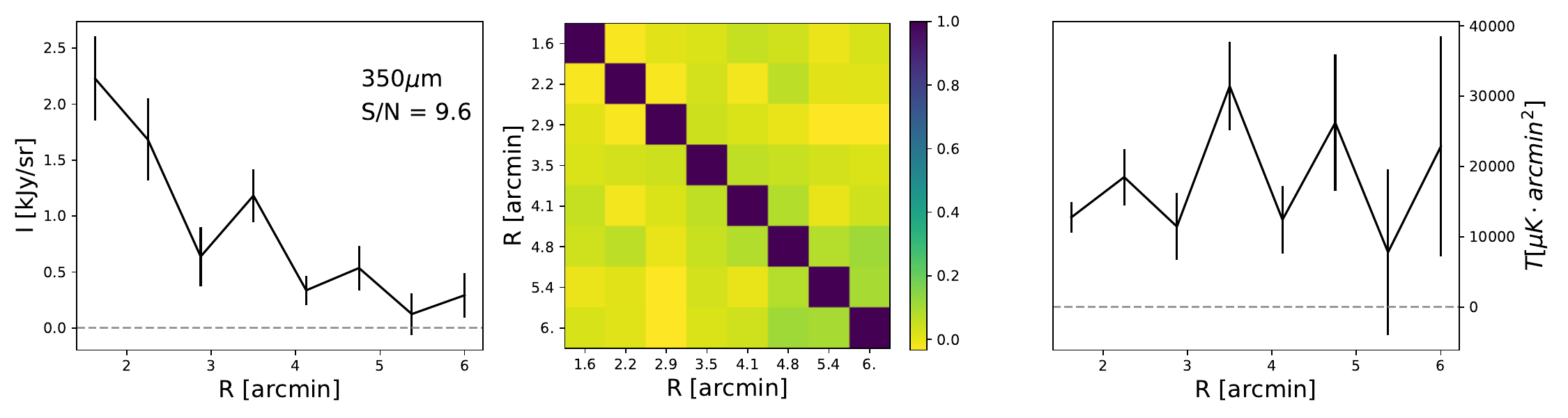}
     \includegraphics[scale=0.3]{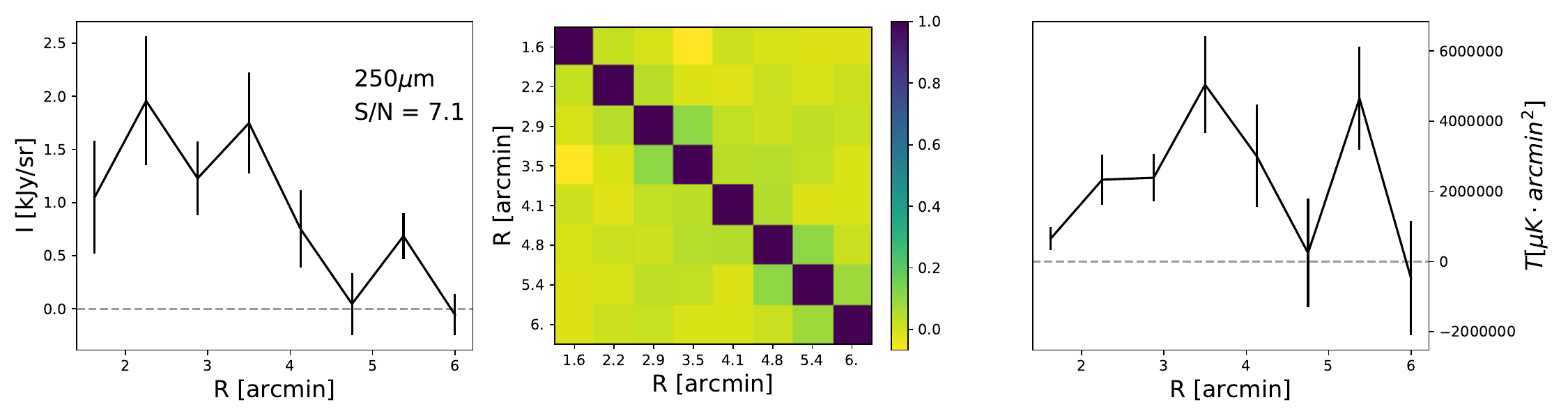}
  \caption{CMASS stacked profiles from H-ATLAS/GAMA.
  For each of the three SPIRE frequency bands, we show the profiles in intensity units $[\rm kJy/sr]$ (left) and cumulative temperature $[\rm \mu K \cdot arcmin^2]$ to match the units of the stacked SZ profiles. 
  The small number density, 0.02 sources/arcmin$^2$ may explain the small covariance among the apertures. We have also tested that the covariance effectively increases at even larger apertures. }
  \label{fig:Iprof}
\end{figure*}

\begin{figure*}
\centering
  \includegraphics[scale=0.3]{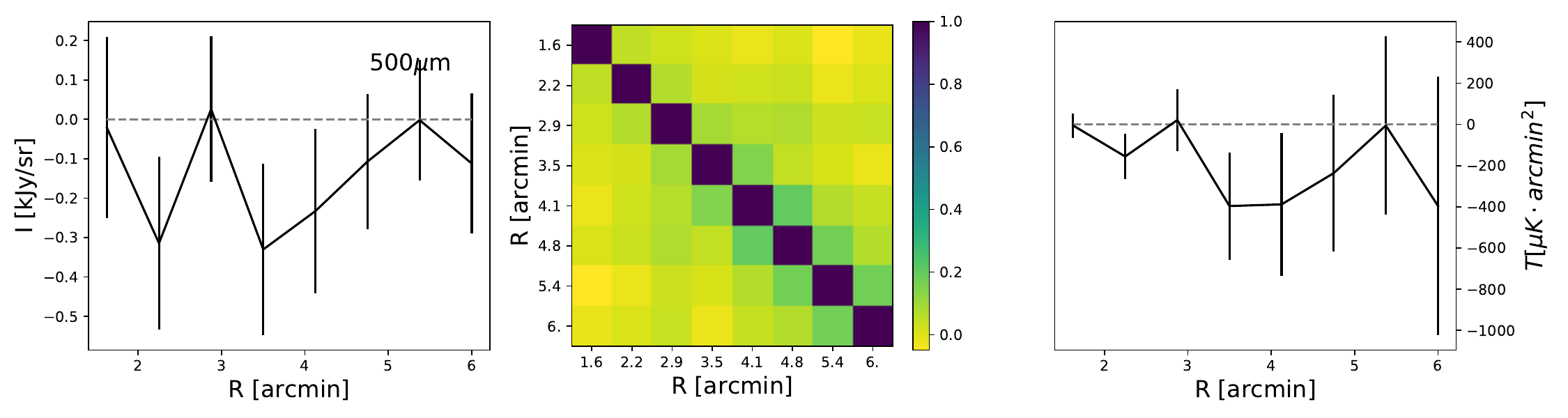}
  \includegraphics[scale=0.3]{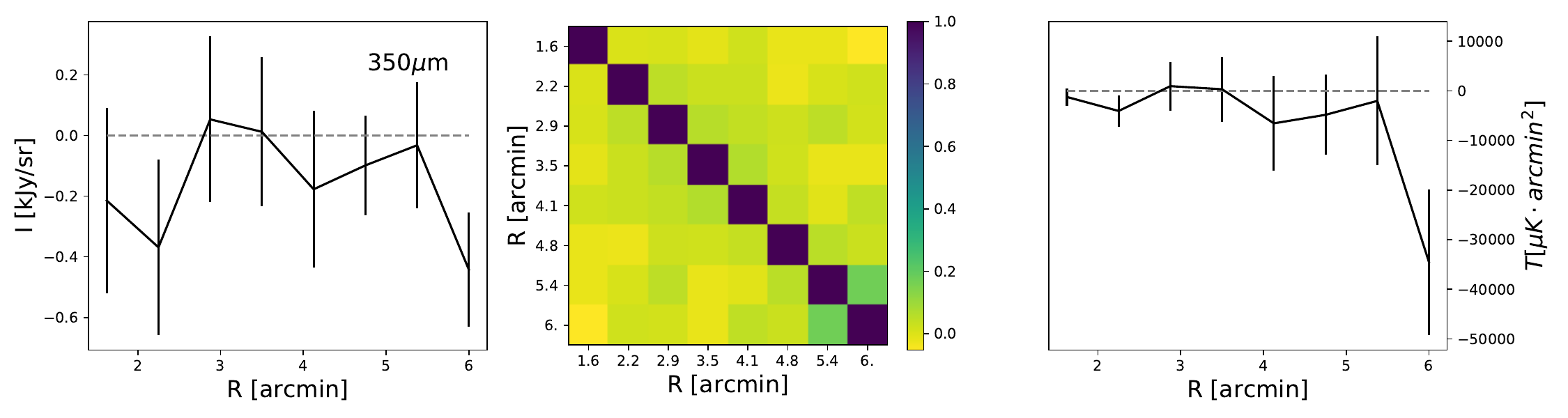}
    \includegraphics[scale=0.3]{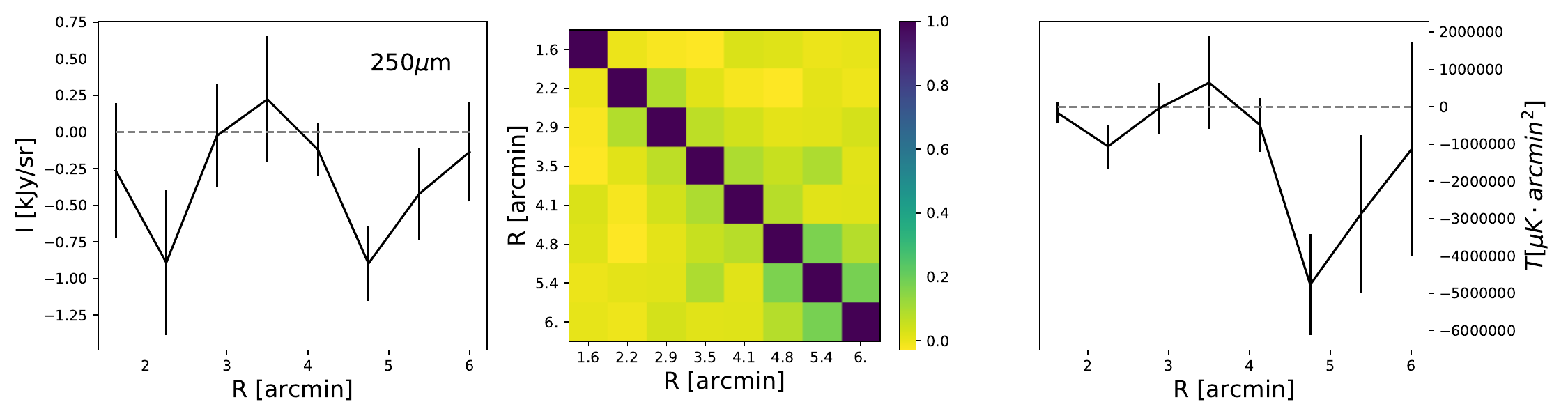}
  \caption{Null tests from stacking on random positions on the maps, for the same number of galaxies in the overlapping area. The $\chi^2$ computed accounting for the covariance in each band is consistent with an average zero signal.
  }
  \label{fig:Irdnprof}
\end{figure*}

A contaminant of the tSZ signal is the light emitted from star-forming CMASS galaxies in the optical/UV that is absorbed by dust grains and re-emitted in the infrared/sub-mm.
This contribution 
must be accounted for in the tSZ signal modeling. On the other hand, since the dust emission does not correlate with the velocities, it does not affect the kSZ measurements.
We model both the frequency and spatial distribution of the dust emission as:
\begin{equation}
\begin{split}
\label{eq:dustmodel}
I(\nu,R) =  &~A_{\rm dust} \left( \frac{\nu (1+z)}{\nu_0} \right)^{\beta_{\rm dust}+3} \frac{e^{(h\nu_0 / k_B T_{\rm dust})}-1}{e^{(h\nu (1+z) / k_B T_{\rm dust})}-1} \\
&\times (c_0 + c_1 R + c_2 R^2)\,,
\end{split}
\end{equation}
where $\nu_0$ is the rest-frame frequency at which we normalize the dust emission, $R$ is the radius of the aperture photometry filter,  $z$ is the redshift of the dust emitters, $A_{\rm dust}$ is the amplitude of the dust emission in [kJy/sr], $\beta_{\rm dust}$ is the  dust spectral index, $T_{\rm dust}$ is the dust temperature in K, and $c_0$, $c_1$, $c_2$ are  the polynomial coefficients parametrizing the radial profile. In order to model the dust in the ACT f090 and f150 bands, we include in our analysis data at larger frequencies where dust emission is dominant over the tSZ. We use Herschel data from one large extragalactic survey that overlaps with ACT, the Herschel Astrophysical TeraHertz Large Area Survey (H-ATLAS \cite{eales+10}), in the three fields GAMA-9, GAMA-12 and GAMA-15 (see Figure \ref{fig:footprint}). 
The H-ATLAS/GAMA survey mapped over 161 deg$^2$ of the sky in five photometric bands: 100 $\mu$m and 160 $\mu$m using the PACS instrument, and 250 $\mu$m, 350 $\mu$m, and 500 $\mu$m using the SPIRE instrument. In this area lie 8871 halos of the ACT+CMASS catalog. 
We use the maps released by the H-ATLAS team in the three SPIRE bands.
We use the raw maps instead of the filtered, background-subtracted maps that are also released because in the latter the signal on scales larger than 3 arcmin has been removed to avoid the contribution from the Milky Way or other large-scale extragalactic emissions, while we are interested in scales up to 6 arcmin that are relevant for feedback effects.
We apply the same aperture photometry and stacking technique used for measuring the tSZ and we obtain the profiles shown in Figure \ref{fig:Iprof}. The results of the null test shown in Figure \ref{fig:Irdnprof} ensure that the measured signal is not a feature of the stacking technique, since stacking on random positions returns a profile consistent with zero on average. The probability to exceed the $\chi^2$ for the null hypothesis is 0.03, 0.41, 0.68 for the profiles at 250 $\mu$m, 350 $\mu$m, and 500 $\mu$m, respectively.
We do not include here data from the Herschel Stripe 82 survey (HerS \cite{viero+14}), also overlapping with our ACT fields. Using HerS maps, we measure a signal that is about 2$\sigma$ smaller than the H-ATLAS/GAMA detection. Since the stacking weights the CAP contributions on the photometric errors, HerS sources do not add significant signal and do not help increasing the S/N, when added to GAMA sources.

\begin{figure*}
\centering
\includegraphics[scale=0.4]{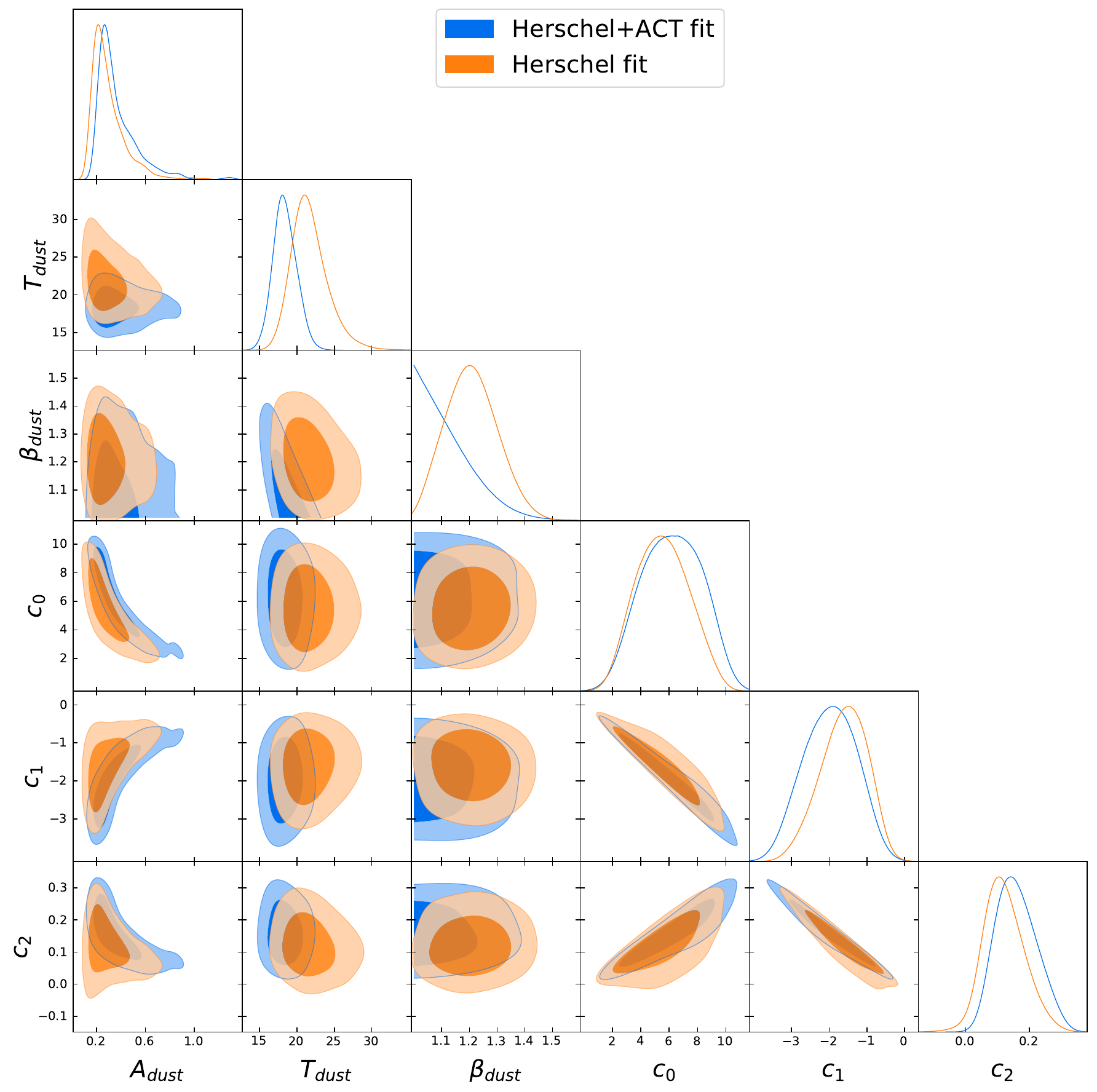} \\
\includegraphics[scale=0.4]{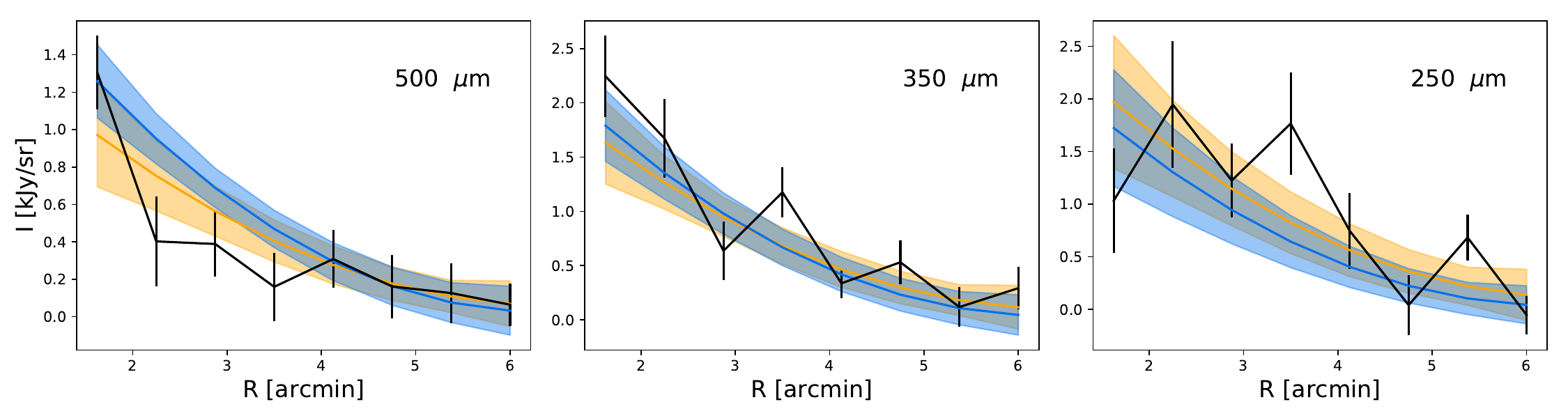} \\
\includegraphics[scale=0.4]{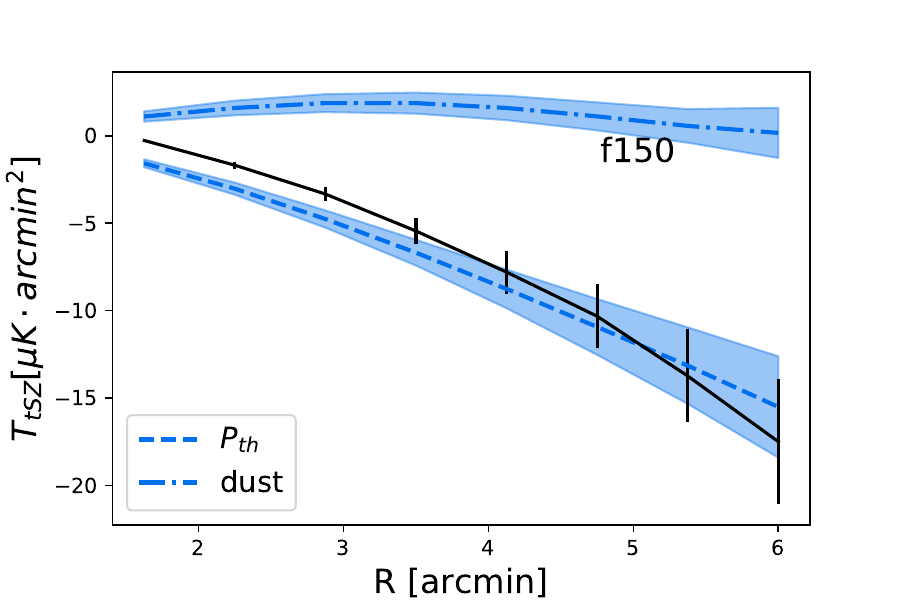} 
\includegraphics[scale=0.4]{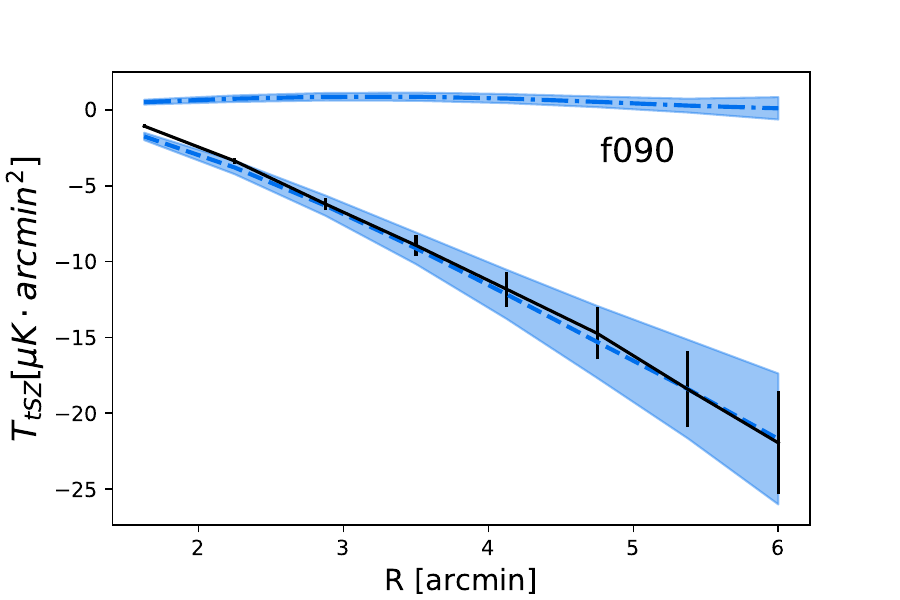} \\
\caption{Top: Constraints on the dust model (\ref{eq:dustmodel}). Results obtained with the simultaneous fit to the GNFW pressure model using ACT and Herschel temperature measurements are shown in blue. Results from the fit of the dust model to Herschel data only are shown in orange. All parameters match within 1$\sigma$. Middle: profiles measured in the three Herschel/SPIRE frequency bands (black) in intensity units [kJy/sr]. The blue and orange curves show the best-fit models to Herschel+ACT and Hershel only data, respectively, and the corresponding bands show the 2$\sigma$ (2nd-98th percentiles) of the distribution of the models obtained from the MCMC chains. Bottom: profiles measured in the two ACT frequency bands (black) in cumulative temperature units of $[\mu K \cdot arcmin^2]$. The blue curves and  2$\sigma$ bands show results obtained with the simultaneous fit of the dust and the GNFW pressure models using Herschel+ACT data. We separate the dust contribution (dot-dashed curves) from the thermal pressure (dashed curves). The best-fit model is the sum of the two contributions. }
\label{fig:dustfit}
\end{figure*}

Figure \ref{fig:dustfit} shows the fit results of the dust model (Eq.\ \ref{eq:dustmodel}) to the Herschel data in orange, and the results of the simultaneous fit of the GNFW pressure model using ACT and Herschel data, in blue. The top panel shows the constraints on the parameters of the dust model obtained in the two cases; all match within $1\sigma$. 
We assume flat priors for the the dust amplitude and temperature parameters in the ranges: $0.05<A_{\rm dust} ~{\rm [kJy/sr]}<5$, $10<T_{\rm dust} ~{\rm [K]}<40$. For the emissivity index we assume a truncated gaussian prior distribution centered on 1.2 and with standard deviation of 0.1, in the range $1<\beta_{\rm dust}<2.5$. These values are consistent with the model used by \cite{Madhavacheril+19} to produce CIB-deprojected y-maps and with the sky-average CIB spectral energy distribution obtained by \emph{Planck} measurements of the CIB power spectra \cite{2014A&A...571A..30P}.  
We also assume flat priors for the polynomial coefficients in the ranges: $0.1<c_0<10$, $-10<c_1<10$, $-10<c_2<10$. 

The middle and bottom panels show the best-fit model over the data in our ACT and Herschel bands, and the $2\sigma$ bounds of the distribution of the models obtained from the MCMC chains.
This analysis is justified by the need to correct our pressure model for dust contamination, which is relevant at 150 GHz (more than at 90 GHz) as shown in the bottom panel of Figure \ref{fig:dustfit}. 
The dust model and the parameters inferred from it are entirely to marginalize over and mitigate the dust contamination in the tSZ signal. These parameters are degenerate with each other and we find consistent values for them throughout our analyses. We make no attempt to infer anything about the dust properties of the CMASS galaxies.

\end{document}